\RequirePackage{rotating}

\documentclass[a4paper,fleqn,usenatbib,useAMS]{mnras}

\usepackage{rotating,graphicx}	% Including figure files
\usepackage{amsmath}	% Advanced maths commands
\usepackage{amssymb}	% Extra maths symbols
\usepackage{multicol}        % Multi-column entries in tables
\usepackage{bm}		% Bold maths symbols, including upright Greek
\usepackage{pdflscape}	% Landscape pages
\usepackage[encapsulated]{CJK}
\usepackage{ucs}
\usepackage[utf8x]{inputenc}
\usepackage[T1]{fontenc}
\usepackage{ae,aecompl}

\usepackage{newtxtext,newtxmath}
\title[PNe with WR-CSPN (I)]{Planetary nebulae with Wolf-Rayet-type central stars - I.\,The case of the high-excitation NGC\,2371}
\author[G\'{o}mez-Gonz\'{a}lez et al.]{V.~M.~A.\,G\'{o}mez-Gonz\'{a}lez$^{1}$\thanks{E-mail:\,v.gomez@irya.unam.mx}, J.~A.\,Toal\'{a}$^{1}$, M.~A.\,Guerrero$^{2}$, H.\,Todt$^{3}$, L.\,Sabin$^{4}$, 
\newauthor{G.\,Ramos-Larios$^{5}$ and Y.~D.\,Mayya$^{6}$}\\
  $^{1}$Instituto de Radioastronom\'{i}a y Astrof\'{i}sica (IRyA), UNAM Campus Morelia, Apartado postal 3-72, 58090 Morelia, Michoac\'{a}n, Mexico\\
  $^{2}$Instituto de Astrof\'{i}sica de Andaluc\'{i}a (IAA-CSIC), Glorieta de la Astronom\'{i}a S/N, 18008 Granada, Spain\\
  $^{3}$Institute for Physics and Astronomy, Universit\"{a}t Potsdam, Karl-Liebknecht-Str. 24/25, D-14476 Potsdam, Germany\\
  $^{4}$Instituto de Astronom\'{i}a, UNAM, Apdo. Postal 877, Ensenada 22860, B.C., Mexico\\
  $^{5}$Instituto de Astronom\'{i}a y Meteorolog\'{i}a, CUCEI, Universidad de Guadalajara, Av. Vallarta 2602, Arcos Vallarta, 44130 Guadalajara, Mexico\\
  $^{6}$Instituto Nacional de Astrof\'{i}sica, \'{O}ptica y Electr\'onica, Luis Enrique Erro 1, Tonantzintla 72840, Puebla, Mexico\\ 
}

\pubyear{2019}
\hypersetup{draft}

\begin{document}
\label{firstpage}
\pagerange{\pageref{firstpage}--\pageref{lastpage}}
\maketitle

\begin{abstract}

We present the analysis of the planetary nebula (PN) NGC\,2371 around the [Wolf-Rayet] ([WR]) star WD\,0722$+$295. 
Our Isaac Newton Telescope (INT) Intermediate Dispersion Spectrograph (IDS) spectra, in conjunction with archival optical and UV images, unveil in unprecedented detail the high-ionisation of NGC\,2371. 
The nebula has an apparent multipolar morphology, with two pairs of lobes 
protruding from a barrel-like central cavity, a pair of dense low-ionisation 
knots misaligned with the symmetry axis embedded within the central cavity, 
and a high excitation halo mainly detected in He\,{\sc ii}. 
The abundances from the barrel-like central cavity and dense knots 
agree with abundance determinations for other PNe with [WR]-type 
CSPNe. 
We suggest that the densest knots inside NGC\,2371 are the oldest structures, 
remnant of a dense equatorial structure, whilst the main nebular shell and 
outer lobes resulted from a latter ejection that ended the stellar evolution. 
The analysis of position-velocity diagrams produced from our high-quality 
spectra suggests that NGC\,2371 has a bipolar shape with each lobe presenting 
a double-structure protruding from a barrel-like central region. 
The analysis of the spectra of WD\,0722$+$295 results in similar 
stellar parameters as previously reported. We corroborate that 
the spectral sub-type corresponds with a [WO1] type.

\end{abstract}


%%%%%%%%%%%%%%%%%%%%%%%%%%%%%%%%%%%%%%%%%%%%%%%%%%

%%%%%%%%%%%%%%%%% BODY OF PAPER %%%%%%%%%%%%%%%%%%

\section{INTRODUCTION}
\label{sec:intro}

Planetary nebulae (PNe) represent a short-lived configuration of the 
circumstellar medium of evolved low- and intermediate-mass stars (1~M$_{\odot} \lesssim M_\mathrm{i} \lesssim$8~M$_{\odot}$).
The interacting stellar winds model of formation predicts that low- and intermediate-mass stars evolve through the asymptotic giant branch (AGB) phase producing a dense and slow wind, which eventually will be swept and compressed by the future fast stellar wind of the post-AGB star and subsequently ionised by the newly developed UV flux \citep[see][]{Kwok1978,Balick1987}. 
The wide variety of PN morphologies \citep[round, bipolar, multipolar, irregular, etc; see][]{Sahai2011} has led to the suggestion that several agents might contribute to the PN shaping, including binarity \citep[or multiple systems;][]{Akashi2017}, jets and magnetic fields \citep[see the review by][]{Zijlstra2015}.

\begin{figure*}
\begin{center}
  \includegraphics[angle=0,width=0.95\linewidth]{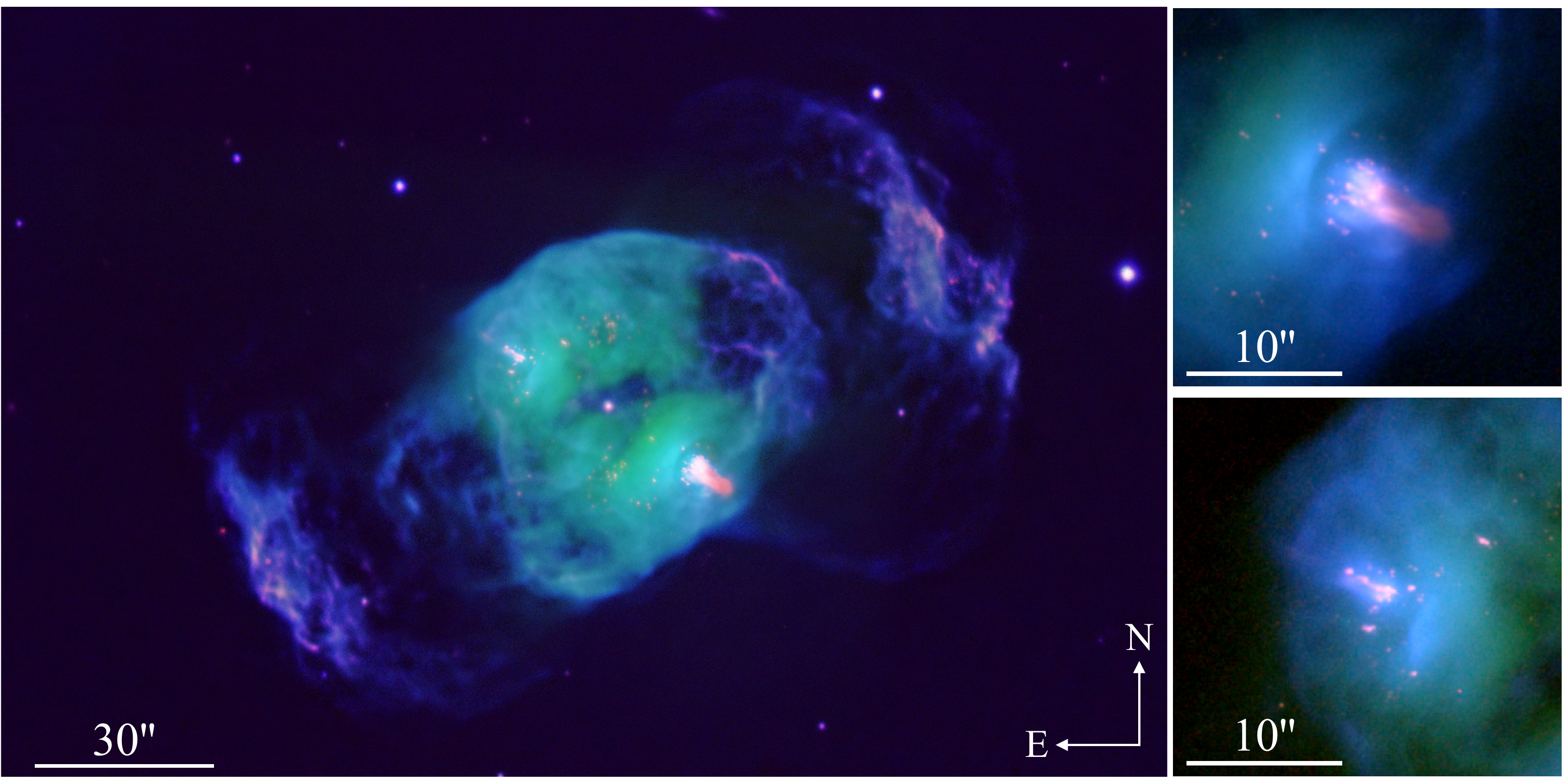}
\label{fig:NGC2371_RGB}
\caption{Colour-composite images of NGC\,2371. The wide-field image
  (left panel) was produced by combining the CFHT and {\it HST}
  observations. The right panels were obtained by only taking the {\it
    HST} observations. In all panels red, green and blue correspond to
  [N\,{\sc ii}], H$\alpha$ and [O\,{\sc iii}], respectively.}
\end{center}
\end{figure*}

Central stars of PNe (CSPNe) commonly show the presence of hydrogen in their surfaces.  
However, about 10\% of CSPNe do not exhibit hydrogen features, i.e., they are hydrogen-deficient, rather displaying emission features in their spectra similar to those of the massive Wolf-Rayet (WR) stars of the carbon sequence \citep[see][and references therein]{Acker2003,Gorny2001}. Accordingly, they are classified in the same way as WR stars, just using square brackets to denote that they are CSPNe instead \citep[][]{Tylenda1993,Crowther1998,Acker2003}.  
It is interesting to note that only a few PNe have been reported to harbor 
hydrogen-poor [WR] CSPN of the nitrogen sequence: the transitional [WN/WC]-type PB\,8
\citep[][]{Todt2010} and the [WN]-type CSPNe of Abell\,48 \citep[][]{Todt2013a} and IC\,4663 \citep[][]{Misza2012}.

In general, abundance determination of PNe with [WR]-type CSPNe do not present major differences compared to those classical PNe with hydrogen-rich CSPNe \citep[e.g.,][]{Pena1998,Pena2001}. 
Hydrogen-deficient material ejected inside the old, hydrogen-rich PN has been identified 
only around a few [WR]-type CSPN \citep[see][and references therein]{Guerrero2018}. These are known as \emph{born-again} PNe and it is accepted that the CSPN experienced a very late thermal pulse (VLTP) ejecting hydrogen-depleted and carbon-rich processed material \citep[e.g.,][]{Herwig1999,MillerBertolami2006}.
Although this evolutionary path has been suggested to explain the formation of
hydrogen-poor [WR] stars, it might not apply to all PNe harboring [WR] stars \citep[see][]{Gorny2000}.

Statistical studies of PNe around [WR] CSPNe have shed light into the relation between the nebular properties and the evolutionary sequence of the progenitor stars. For example, the decrease in electron density ($n_\mathrm{e}$) with decreasing spectral subtype has been suggested to be a signature of an evolutionary sequence from the late to early [WC]-type \citep{Acker1996,Gorny2000}. Furthermore, the N/O abundance ratio vs.\ [WC] class presented by \citet{Pena2001} confirmed the claim that stars with different masses evolve through the same [WC] state 
\citep[e.g.,][]{Pena1998,DeMarco1999}. 

Comparisons between PNe harboring non-emission line CSPNe with those hosting [WR] CSPNe 
have unveiled subtle differences. 
PNe with [WR] CSPNe are generally more centrally condensed, expand faster, and 
exhibit a higher degree of turbulence than PNe with non-emission line CSPNe 
\citep[e.g.,][and references  therein]{Gesicki2006,Medina2006,Jacob2013}. 
  They also seem to have higher N and C abundances, implying that they have evolved 
  from stars with initial masses $\sim$4~M$_{\odot}$ \citep{GarciaRojas2013}.  
  This is consistent with their preferential location at low latitudes 
  in the Galactic disk \citep{Pena2013}.  
  Finally, it has been noted recently that PNe with [WC] CSPNe have a higher 
  occurrence of fast collimated outflows than PNe with weak emission line CSPNe or non-emission line CSPNe \citep{Rechy2017}.

These statistical studies rely on averaged values of the nebular physical properties 
(electron density and temperature), abundances, and velocity fields
that might hide important clues on the evolution of PNe with [WR]-type CSPNe. 
  For example, dedicated studies of compact [WR] PNe have unveiled the effects of 
  fast collimated outflows in their early shaping \citep[][]{Rechy2017,Rechy2019}.
  Detailed studies of individual PNe with [WR] CSPNe are clearly needed to understand this unique evolutionary path of low- and intermediate-mass stars. We have therefore started a series of detailed studies of PNe harboring [WR] CSPNe. 
In this first paper we present a study of NGC\,2371 
\citep[a.k.a. PNG\,189.1+19.8;][]{Curtis1918}, which 
harbors the [WR] CSPN WD\,0722$+$295.

NGC\,2371 has been classified as a barrel-like PNe of Peimbert type II \citep[see table~1 in][]{Henry2018}.  
Morphological studies performed with the \emph{Hubble Space Telescope} (\emph{HST}), \emph{Spitzer} and ground-based telescopes have unveiled a variety of morphological features \citep[see Fig.~1;][]{Sabbadin1982,RamosLarios2012}:
two-bipolar lobes extending $\gtrsim$1~arcmin from the CSPN towards the NW and SE directions with position angle PA$\approx -60^{\circ}$, a main central cavity with an ellipsoidal shape with an extension of 28$^{\prime\prime}\times$38$^{\prime\prime}$ that exhibits a couple of blowout features aligned with the outer lobes, and a collection of dense knots mainly detected in
[N\,{\sc ii}] and [S\,{\sc ii}] narrow-band filter images (see next section).
The two brightest knots are aligned in the NE to SW direction with a PA$\approx60^{\circ}$, but the line that connects them is not perfectly aligned with the CSPN. \citet{RamosLarios2012} presented a detailed discussion on the properties of the dense NE and SW knots. Their \emph{HST} images show that these are actually conglomerates of clumps (see also Fig.~1). 
A secondary pair of smaller size clumps is observed at a PA=5$^\circ$. 
It is worth noting that the low-ionisation clumps do not have a uniform 
distribution around the CSPN nor are they aligned with the bipolar lobes.

Optical \citep{TorresPeimbert1977,Aller1979} and \emph{International Ultraviolet Explorer} (\emph{IUE}) UV \citep{Pottasch1981} observations of NGC\,2371 have been used to estimate an averaged electron temperatures ($T_\mathrm{e}$) in the range 9,000--23,000~K, depending on the line ratio used or the adopted electron density ($n_\mathrm{e}$), which is found to be 800--2,500~cm$^{-3}$ \citep[e.g.,][]{Pena2001}.
\citet{Kaler1993} presented the analysis of the optical spectrum in the 
blue region ($\sim$3700--5000~\AA) and showed it to be dominated by the 
O\,{\sc vi}~$\lambda\lambda$3811,3834 doublet with contributions from the
He\,{\sc ii}~$\lambda$4686, C\,{\sc iii}~$\lambda$4650 and 
C\,{\sc iv}~$\lambda$4658. 
These authors estimated an effective temperature ($T_\mathrm{eff}$) for  
WD\,0722$+$295 of about $T_\mathrm{eff}$=[1--1.2]$\times$10$^{5}$~K, 
consistent with the analysis of \emph{IUE} observations \citep{Pottasch1981}.
  Since its early classification as a [WR] star of the oxygen sequence \citep{Smith1969}, 
  its spectral type has swung from early [WC3] \citep[][]{Heap1982} to [WO1] \citep{Acker2003}.

In this work, we intend to unveil the physical structure of the different components 
of NGC\,2371 and to produce an accurate model of its central star. 
%In this paper we present an analysis of the PN NGC\,2371. 
We have obtained Isaac Newton Telescope (INT) spectroscopic observations to study the
physical properties (density and temperature) and abundances to
describe the ionisation structure of the different morphological
components of NGC\,2371. 
In conjunction with archival \emph{IUE} observations, the INT spectra are also used to perform a detailed characterisation of its CSPN. 
The paper is organized as follows. 
In Section~2 we describe the observations used in the present work. 
Section~3 presents the analysis and atmospheric model of WD\,0722$+$295. 
In Section~4 we present our spectral analysis of NGC\,2371 as well as the estimates of the physical structure. 
In Section~5 we discuss our results and, finally, a summary is presented in Section~6.

\section{Observations}
\label{sec:obs}

\begin{figure}
\begin{center}
  \includegraphics[angle=0,width=0.95\linewidth]{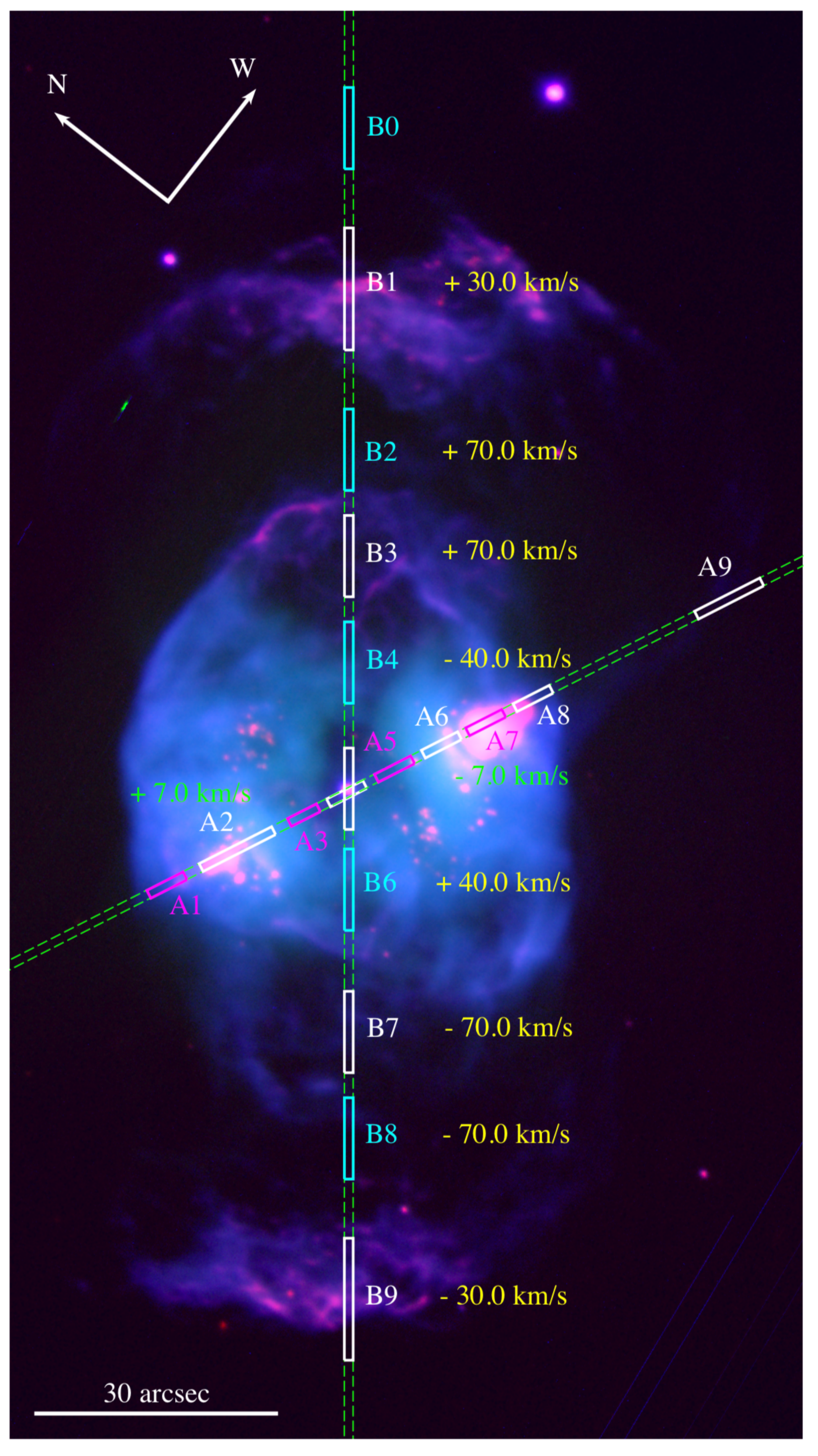}
\label{fig:NGC2371_slits}
\caption{CFHT colour-composite ([O\,{\sc iii}] - blue, H$\alpha$ -   green, [N\,{\sc ii}] - red) image of NGC\,2371. The (green) dashed-line regions represent the slits positions obtained from our INT IDS observations. The PA for the slits are 65$^{\circ}$ (Slit\,A - minor axis) and 128$^{\circ}$ (Slit\,B - major axis). Different extraction regions are labeled and shown with different colours. The green and yellow velocity measurements correspond to the A2, A7 and
B1-B9 extraction regions. See text for details.}
\end{center}
\end{figure}

Narrow-band images of NGC\,2371 were downloaded from the archives of the Canada-France-Hawaii Telescope (CFHT)\footnote{\url{http://www.cadc-ccda.hia-iha.nrc-cnrc.gc.ca/en/}} and \emph{HST} Legacy Archive\footnote{\url{https://hla.stsci.edu/}}. The CFHT images were acquired with the multi-object spectrograph (MOS) camera. 
These images were obtained through the H$\alpha$, [O\,{\sc iii}] and [N\,{\sc ii}] filters with total exposure times of 1200~s on each filter on December 18, 2002 (Prop.\,ID.: 02BC17; PI: S.\,Kwok). 
The {\it HST} images were obtained through the F502N, F656N, and F658N filters (hereafter [O\,{\sc iii}], H$\alpha$ and [N\,{\sc ii}]) on 2007 November 15 with exposure times on each filter of 1600 s (Prop.ID.\,11093, PI: K.\,Noll). A colour-composite image obtained by combining all these optical images is shown in Figure~1.

Long-slit spectroscopic observations were obtained at the INT of the Observatorio de El Roque de los Muchachos (La Palma, Spain) using the Intermediate Dispersion Spectrograph (IDS) in low- and high-dispersion spectral modes (PI: M.~A.\,Guerrero). 
The high-resolution spectra were obtained on 2018 October 24 
using the EEV10 CCD camera with the R1200U grating.  
This configuration has a dispersion of 0.48~\AA~pixel$^{-1}$ over the 3200--4500 \AA\ 
wavelength range and a plate scale of 0\farcs44~pixel$^{-1}$. 
The low-resolution spectra were acquired on 2018 November 15 
using the RED$+$2 CCD camera with the R400V grating. 
This configuration has a dispersion of 1.55~\AA~pixel$^{-1}$ over the 3500--8000 \AA\ 
wavelength range and a similar plate scale of 0\farcs44~pixel$^{-1}$.
The slit width was set at 1\farcs1 in both observing runs.
Two slits at PA=65$^\circ$ (Slit~A) and PA=128$^\circ$ (Slit~B) were used with both spectral configurations. Slit B is centered on WD\,0722$+$295, whereas Slit A probes the low-ionisation knots and therefore it does not goes exactly across the CSPN. The positions of these slits are overlaid on the colour-composite image of NGC\,2371 in Figure~2. The total integration time on each position and spectral configuration was 900~s.  
All spectra were analysed following {\sc iraf} standard routines.  
The wavelength calibration was performed using CuAr+CuNe lamps.

\begin{figure*}
\begin{center}
\includegraphics[angle=0,width=1.0\linewidth]{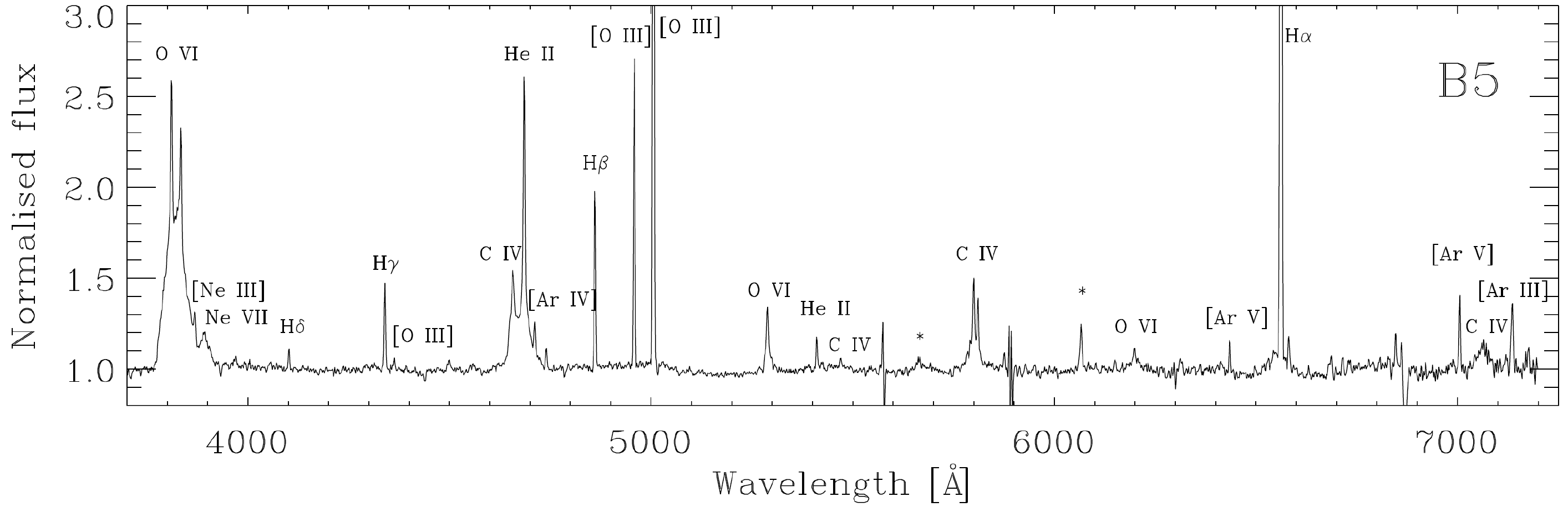}~
\label{fig:NGC2371_spec_star}
\caption{INT IDS low-resolution, normalised spectrum of the CSPN of
  NGC\,2371 obtained from slit\,B. The most prominent WR features and nebular lines are marked.}
\end{center}
\end{figure*}

In order to study the ionisation structure of NGC\,2371, we have
extracted spectra from different regions defined on each slit
  using the {\sc iraf} task {\it apall}.
Those extracted from Slit~A were used to study the central main cavity of 
NGC\,2371 and the dense knots located $\gtrsim$16$^{\prime\prime}$ from
the CSPN, and those from Slit~B to study the outermost lobes and different 
regions of the central cavity along the NW-SE direction.  
All regions, labeled from A1--A9 and B0--B9, are shown in Figure~2 
  and their sizes are indicated in Tables 3 and 4, respectively.
We first extracted the 1-D spectrum of the CSPN (regions A4 and B5) tracing the 
stellar continuum along the 2-D spectrum.  
This information was then used as a reference to trace the nebular spectra (regions A1--A3, 
A5--A9, B0--B4, and B6--B9) along the 2-D spectra, as these regions does not show continuum emission.
Since Slit~A misses some stellar flux, the spectrum extracted for region A4 
has not been used for stellar modelling.
Finally, we note that all extracted spectra were corrected for extinction by
using the $c$(H$\beta$) value estimated from the Balmer decrement method.
We assume intrinsic Balmer decrement ratio corresponding to a case B 
photoionised nebula of $T_{\rm e}$ = 10,000~K and $n_{\rm e}$ = 100~cm$^{-3}$ \citep[see][]{Osterbrock2006} and the reddening curve of \cite{Cardelli1989}.

\section{WD\,0722$+$295 - the CSPN of NGC\,2371}

The optical spectrum of WD\,0722$+$295 obtained from our INT IDS observations 
is presented in Figure~3.  
The stellar spectrum exhibits the classic WR blue and red bumps (BB and RB) at 
$\sim$4686~\AA\ and $\sim$5806~\AA, respectively, as well as a broad O\,{\sc vi} 
feature at $\sim$3820~\AA\ \citep[see also figure~1 in][]{Kaler1993}. 
Other emission lines related to WR features such as those reported in 
\citet{Acker2003} -- namely, O\,{\sc vi} 5290~\AA, C\,{\sc iv} 5470~ \AA\ 
and 7060~\AA, and He\,{\sc ii} 5412~\AA\ -- are also present in the optical spectrum. 
Furthermore, we also detect the O\,{\sc vi} 6200~\AA\, as a broad emission line. 
All identified WR features as well as some contributing nebular lines are marked in Figure~3.
We note the presence of narrow emission lines at $\approx$5665~\AA\ and 
$\approx$6066~\AA\ marked with an asterisk in Figure~3 and Figure~4.  
There has been controversy in the identification of these lines, which can be 
assigned either to the O\,{\sc vii} $\lambda$5666 and O\,{\sc viii} 
$\lambda$6068 lines or to the Ne\,{\sc vii} $\lambda$5666 and Ne\,{\sc viii} 
$\lambda$6068 lines.  
This is discussed in Section 5.

\begin{figure*}
\begin{center}
\includegraphics[angle=0,width=\linewidth]{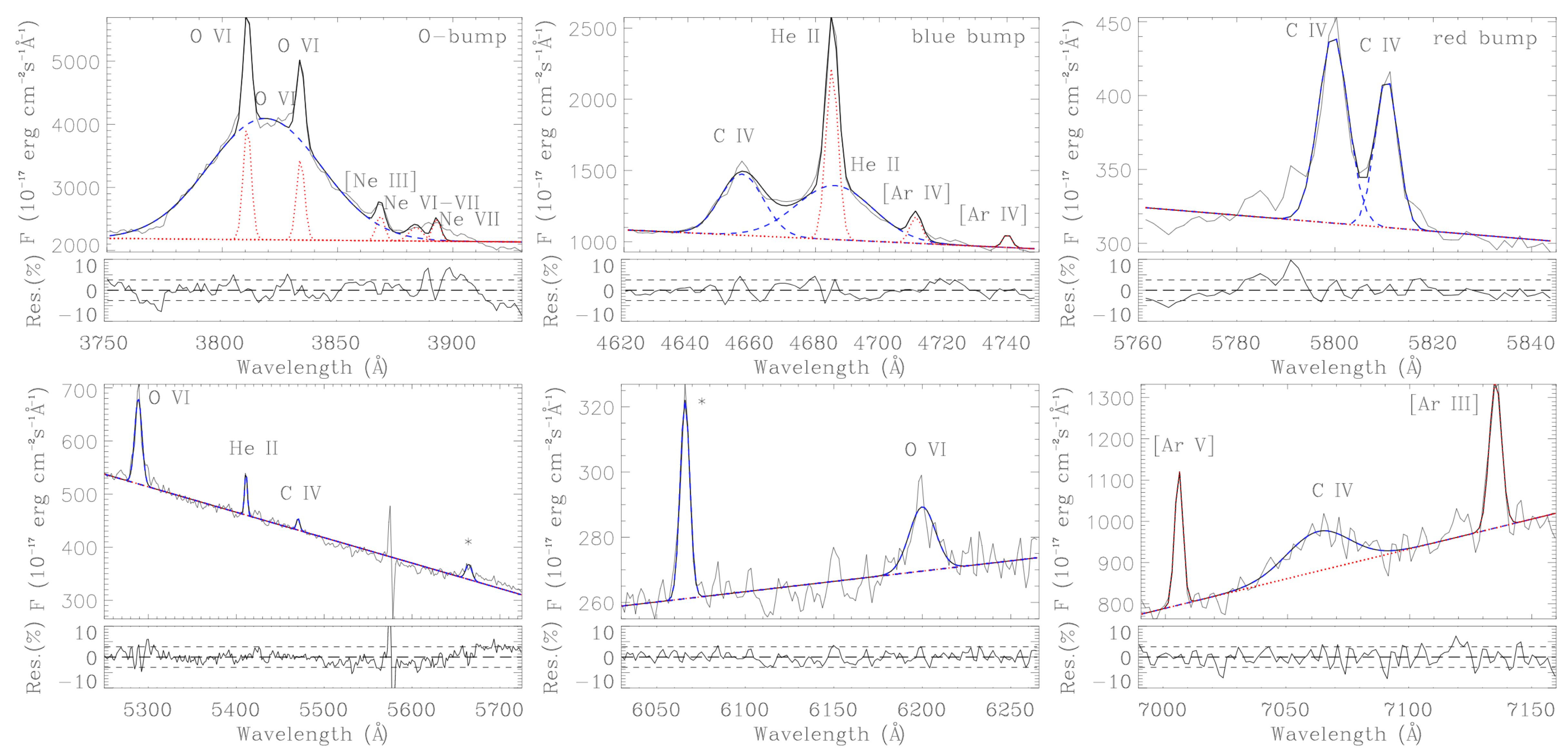}
\label{fig:NGC2371_spec_star_2}
\caption{Multi-component Gaussian fits to the O-bump ({\it top left}), blue bump ({\it top center}) and red bump ({\it top right}) panels. The bottom panels show other WR features in the spectrum of WD\,0722$+$295. The (blue) dashed lines represents the fits to the broad (stellar) features, the (red) dotted-line the fits to the nebular emission and the (black) line the sum. Residuals are shown at the bottom of each panel.}
\end{center}
\end{figure*}

To provide an appropriate description of the spectrum and spectral classification of WD\,0722$+$295, the contribution from nebular lines blended with the broad WR features needs to be accounted for. This was done by applying the analysis described by \citet{GomezGonzalez2020}. This method consists on fitting the broad WR features with multi-Gaussian components using a tailor-made code that uses the {\sc idl} routine {\sc lmfit}\footnote{The {\sc lmfit} function (lmfit.pro) performs a non-linear least squares fit to a function with an arbitrary number of parameters. It uses the Levenberg-Marquardt algorithm, incorporated in the routine {\it mrqmin} from \citet{Press1992}.}. As a result, fluxes, central wavelength, FWHM and equivalent widths (EW) from the WR spectral features as well as contributing nebular lines have been estimated. All these parameters are listed in Table~1. We note that those lines listed as WR features have FWHM larger than those of the nebular narrow lines, that is, FWHM$>4.5$~\AA.

\begin{table}
\begin{center}
\caption{\label{tab:elines} Parameters for the emission lines considered for the
  multi-Gaussian fitting of the WR features of the central star in NGC\,2371.}
\setlength{\tabcolsep}{0.6\tabcolsep}
\begin{tabular}{cllrrrrrr}
\hline
ID & & 
\multicolumn{1}{c}{Ion} & 
\multicolumn{1}{c}{$\lambda_0$} & 
\multicolumn{1}{c}{F} & 
\multicolumn{1}{c}{FWHM} & 
\multicolumn{1}{c}{EW} & 
\multicolumn{1}{c}{$\lambda_\mathrm{c}$} & 
\multicolumn{1}{c}{$F/F_\mathrm{RB}$} \\
\multicolumn{1}{c}{(1)} & 
\multicolumn{1}{c}{(2)} & 
\multicolumn{1}{c}{(3)} & 
\multicolumn{1}{c}{(4)} & 
\multicolumn{1}{c}{(5)} & 
\multicolumn{1}{c}{(6)} & 
\multicolumn{1}{c}{(7)} & 
\multicolumn{1}{c}{(8)} & 
\multicolumn{1}{c}{(9)} \\
\hline
BB         &    &              &             &      &      &      &        &     \\
1          & WR & He\,{\sc ii} & 4686        & 11.4 &28.2  &11.2  & 4686.0 & 712.5 \\
2          & WR & C\,{\sc iv}  & 4658        &  7.3 &15.9  & 7.0  & 4656.9 & 456.3 \\
3          & Neb& He\,{\sc ii} & 4686        &  5.8 & 4.5  & 5.7  & 4685.2 & $\dots$~~  \\
4          & Neb& Ar\,{\sc iv} & 4711        &  0.9 & 4.5  & 0.9  & 4711.5 & $\dots$~~  \\
RB         &    &              &             &      &      &      &        &     \\
5          & WR & C\,{\sc iv}  & 5801        &  1.0 &  7.1 & 3.1  & 5799.7 & $\dots$~~  \\
6          & WR & C\,{\sc iv}  & 5812        &  0.6 & 5.4  & 1.9  & 5810.5 & $\dots$~~  \\
OB         &    &              &             &      &      &      &        &     \\
7          & WR & O\,{\sc vi}  & 3820        &117.6 &57.5  &53.2  & 3818.7 & 7350.0 \\
8          & Neb& O\,{\sc vi}  & 3811        &  8.0 & 4.2  & 3.6  & 3810.9 & $\dots$~~  \\
9          & Neb& O\,{\sc vi}  & 3834        &  5.7 & 4.1  & 2.6  & 3834.0 & $\dots$~~  \\
other      &    &              &             &      &      &      &        &     \\
10         & WR & O\,{\sc vi}  & 5290        &  1.7 & 9.6  & 3.2  & 5288.7 & 106.3 \\
11         & Neb& He\,{\sc ii} & 5412        &  0.3 & 3.6  & 0.7  & 5411.1 & $\dots$~~  \\
12         & WR & C\,{\sc iv}  & 5470        &  0.1 & 4.6  & 0.3  & 5470.4 & 6.3 \\
13 & WR & Ne\,{\sc vii}$^\dagger$& 5666 &  0.3 & 8.2  & 0.8  & 5664.6 & 18.8 \\
14 & WR & Ne\,{\sc viii}$^\dagger$& 6068&  0.4 & 6.4  & 1.6  & 6066.0 & 25.0 \\
15         & WR & O\,{\sc vi}  & 6200        &  0.4 &18.0  & 1.4  & 6200.0 & 25.0 \\
16         & WR & C\,{\sc iv}  & 7060        &  3.1 &30.3  & 3.5  & 7062.5 & 193.8 \\
\hline
\end{tabular}\\
(1) Identification number of the Gaussian components in the WR features;
(2) Nature of the contributing emission line: WR (broad) or nebular (narrow);
(3) Ion responsible for the line;
(4) Rest wavelength in \AA;
(5) Flux in units of $10^{-14}$~erg~cm$^{-2}$~s$^{-1}$;
(6) Full Width at Half Maximum (FWHM) [\AA];
(7) Equivalent Width (EW) [\AA];
(8) Observed center of the line;
(8) Line fluxes normalised to the RB. The ratio has been computed adopting a $F$(RB)=100.
$^\dagger$This identification is discussed in Section~5.
\end{center}
\end{table}

Figure~4 shows the fits to the different WR features of the WD\,0722$+$295. Two of these bumps are composed of several blended emission lines and need a careful modelling in order to separate their broad and narrow contributions. The O-bump feature is made of a broad O\,{\sc vi} (FWHM of 57.5~\AA), two O\,{\sc vi} narrow lines (FWHM$<$4.5~\AA) and several Ne lines at the red wing (see Table 1). 
It is clear that the broad O\,{\sc vi}~3820~\AA\ has a stellar origin and, although the other two O\,{\sc vi} lines have FWHM that suggest a nebular origin, some contribution from the star is expected according to our stellar atmosphere model (see Section~3.1 and Fig.~5). The Gaussian-fitting method shows that the O\,{\sc vi} lines at 3811~\AA\ and 3834~\AA\ contribute about 10\% to the O-bump. Thus, this can be considered as a upper limit to the nebular contribution. Other narrow emission lines also contribute to the BB. For example, the He\,{\sc ii}, which is blended with WR features contributes $\simeq$30\% of the total flux of the BB (He\,{\sc ii}~4686~\AA$+$C\,{\sc iv}~4658~\AA; see Table~2). There is no contribution of nebular lines to the RB (C\,{\sc iv}~5801~\AA\ + C\,{\sc iv}~5812~\AA).

With reliable flux estimates for each WR feature, we can assess the spectral 
classification of WD\,0722$+$295.  
For this, we estimated the line ratios of oxygen, carbon and helium WR features 
relative to the RB (Column~9 of Table~1) and compared them to those defining  
  the quantitative classification scheme of [WR] stars listed in table~2 of 
  \citet{Acker2003}. According to this scheme, WD\,0722$+$295 fulfills the criteria for a [WO1]-type star, but we note that the FWHM of the RB is around half of that suggested for this spectral type. One could argue that using dereddened flux ratios is model dependent and less accurate than using EW ratios. For this, we also compared our results listed in Table~1 with the classification scheme proposed in \citet{Crowther1998} and found that the EW ratio of the O\,{\sc vi}~3820~\AA\ WR feature over the RB is 73.5, above the limit of $\sim$1.6 suggested by these authors for a [WO1] spectral type.

\subsection{NLTE analysis of WD\,0722$+$295}

To obtain a more quantitative description of the properties of WD\,0722$+$295, we have analysed optical and UV spectra by means of the updated version of the Potsdam Wolf-Rayet (PoWR)\footnote{\url{http://www.astro.physik.uni-potsdam.de/~wrh/PoWR}} NLTE stellar atmosphere code \citep[][]{Grafener2002,Hamann2004}. Details of the computing scheme can be found in \citet{Todt2015} and the most recent 
example of the capabilities of the PoWR code performed by our group can be found in  \citet{Toala2019} for the case of the [WR]-type CSPN of NGC\,40.

Available UV \emph{Far Ultraviolet Spectroscopic Explorer} (\emph{FUSE}) and \emph{IUE} observations of WD\,0722$+$295 were retrieved from the Mikulski Archive for Space Telescopes\footnote{\url{https://archive.stsci.edu/hst/}}. The UV observations were analysed in conjunction with the optical INT IDS spectra with the PoWR stellar atmosphere code. The \emph{FUSE} observations correspond to Obs.\,ID. p1330301000 (PI: L.\ Bianchi) and were obtained on 2000 February 26 with the LWRS aperture for a total exposure time of 5259~s. The {\it IUE} data, taken on 1979 April 7, correspond to the Obs.\,ID. SWP04883 and LRW04210 (Program ID SP127; PI: S.\,R.\,Pottasch) with exposure times of 1560~s and 2700~s, respectively. The spectra imply a reddening of $E(B-V)=0.08$~mag using the Cardelli extinction law which is consistent with the value estimated by \citet[][]{Herald2004}.

\begin{table}
  \caption{Parameters of WD\,0722$+$295 obtained with PoWR.}
  \label{tab:parameters}
\begin{tabular}{lcl}
\hline
Parameter & Value & Comment\\
\hline
$T_\mathrm{eff}$ [kK]         &  130    & Defined at $\tau_\text{Ross}=20$ \\
$d$ [kpc]                   &    1.75   &   \citet{bailerjones2018}\\
$\log(L_{\star}/L_\odot)$   &    3.45   &  \\
$R_{\star}$ [$R_\odot$]     &    0.105  & \\
%Suggested by Helge:
$R_\text{t}$ [$R_\odot$]   & 20\\
$D$                         &   10    & Density contrast\\
$\log(\dot{M}/\mathrm{M}_\odot\,\text{yr}^{-1})$ & $\mathrm{-7.75}$ & for $D$=10 \\
$v_{\infty}$ [km\,s$^{-1}$] & 3700    & \\
$M_{\star}$ [M$_\odot$]    & 0.6 & Stellar mass adopted \\
\hline
\multicolumn{3}{l}{Chemical abundances (mass fraction)}\\
\hline
He & 0.71  & \\
C  & 0.20  & \\
N  & 0.001 & \\
O  & 0.06  & \\
Ne & 0.03  & \\
Fe & 1.4$\times10^{-3}$ & \\
\hline
\end{tabular}
\end{table}

\begin{figure*}
\begin{center}
\includegraphics[angle=0,width=0.85\linewidth]{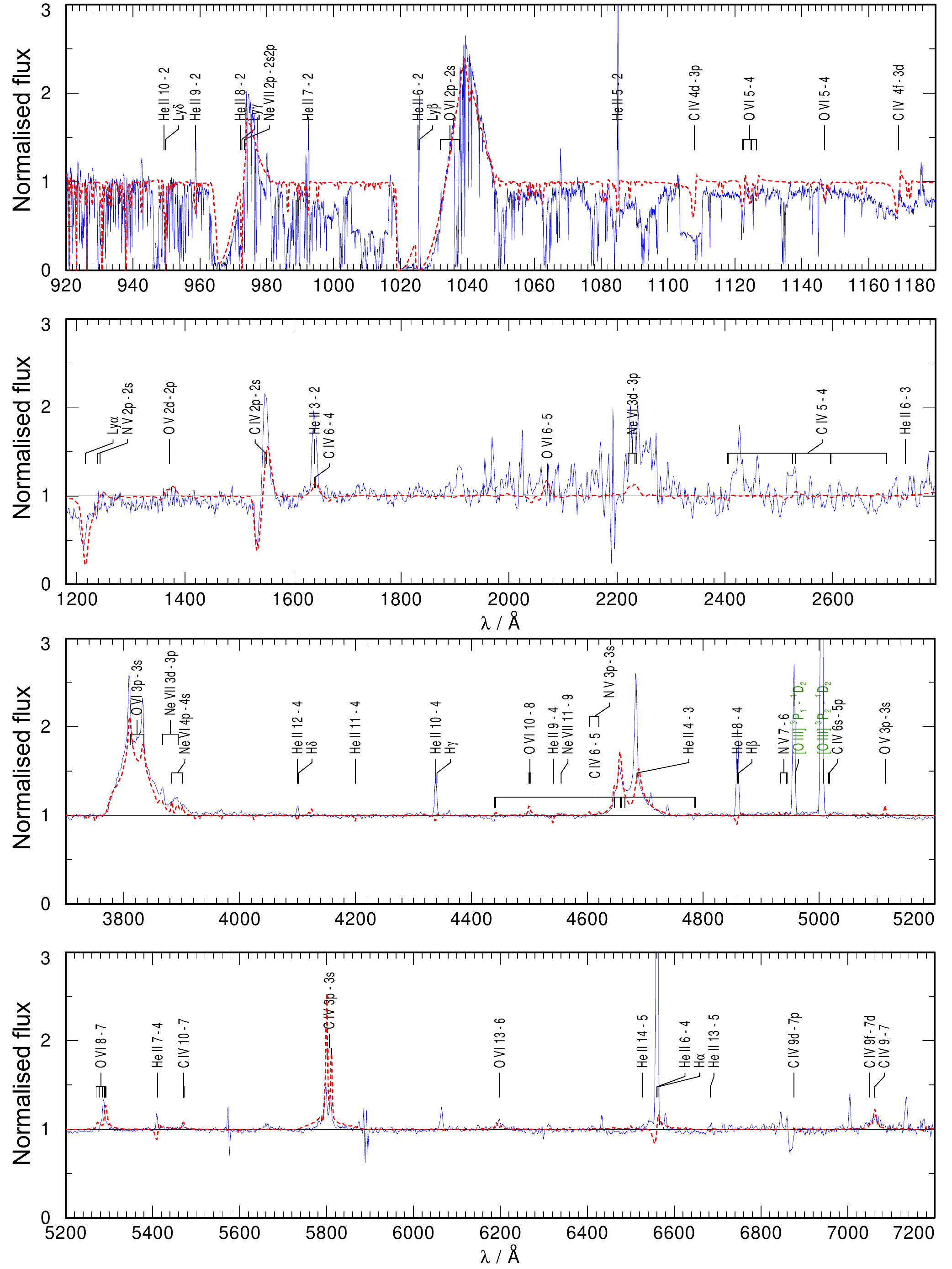}
\label{fig:PoWR}
\caption{Comparison between our PoWR model (red dashed line) with optical and UV observations (blue solid line).}
\end{center}
\end{figure*}

\begin{figure*}
\begin{center}
\includegraphics[angle=0,width=0.85\linewidth]{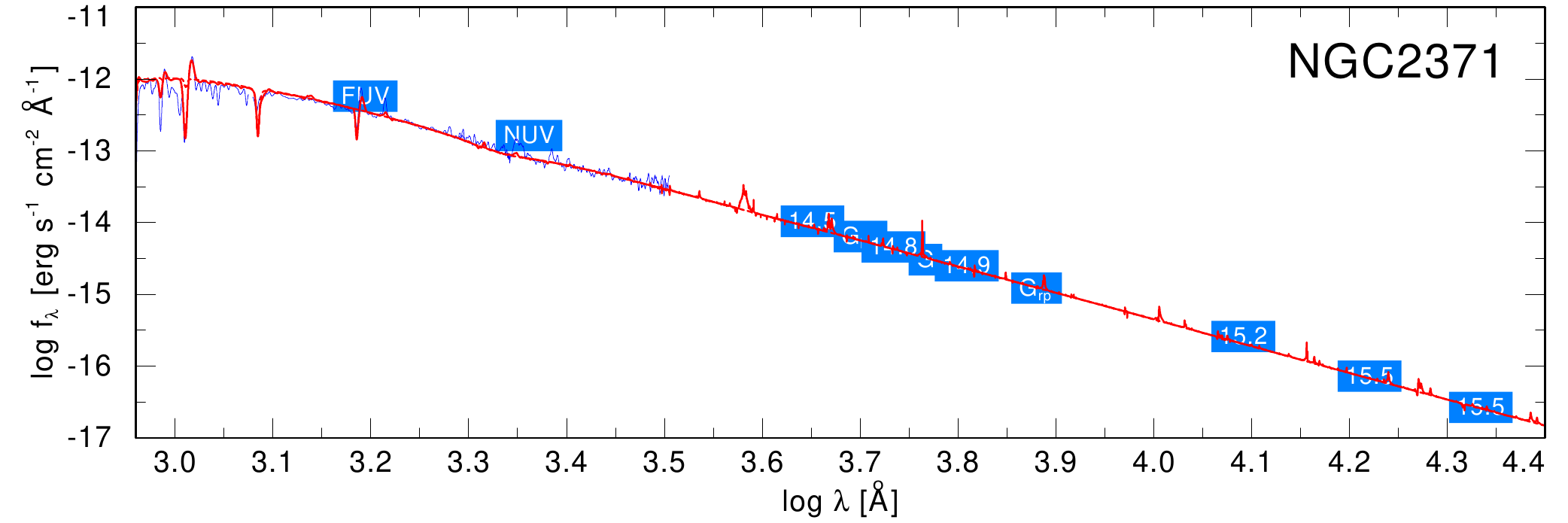}
\label{fig:PoWR_2}
\caption{Comparison between our PoWR model (red line) with observed spectra in the optical and UV and IR photometry (blue).}
\end{center}
\end{figure*}

\begin{figure*}
\begin{center}
\includegraphics[angle=0,width=0.85\linewidth]{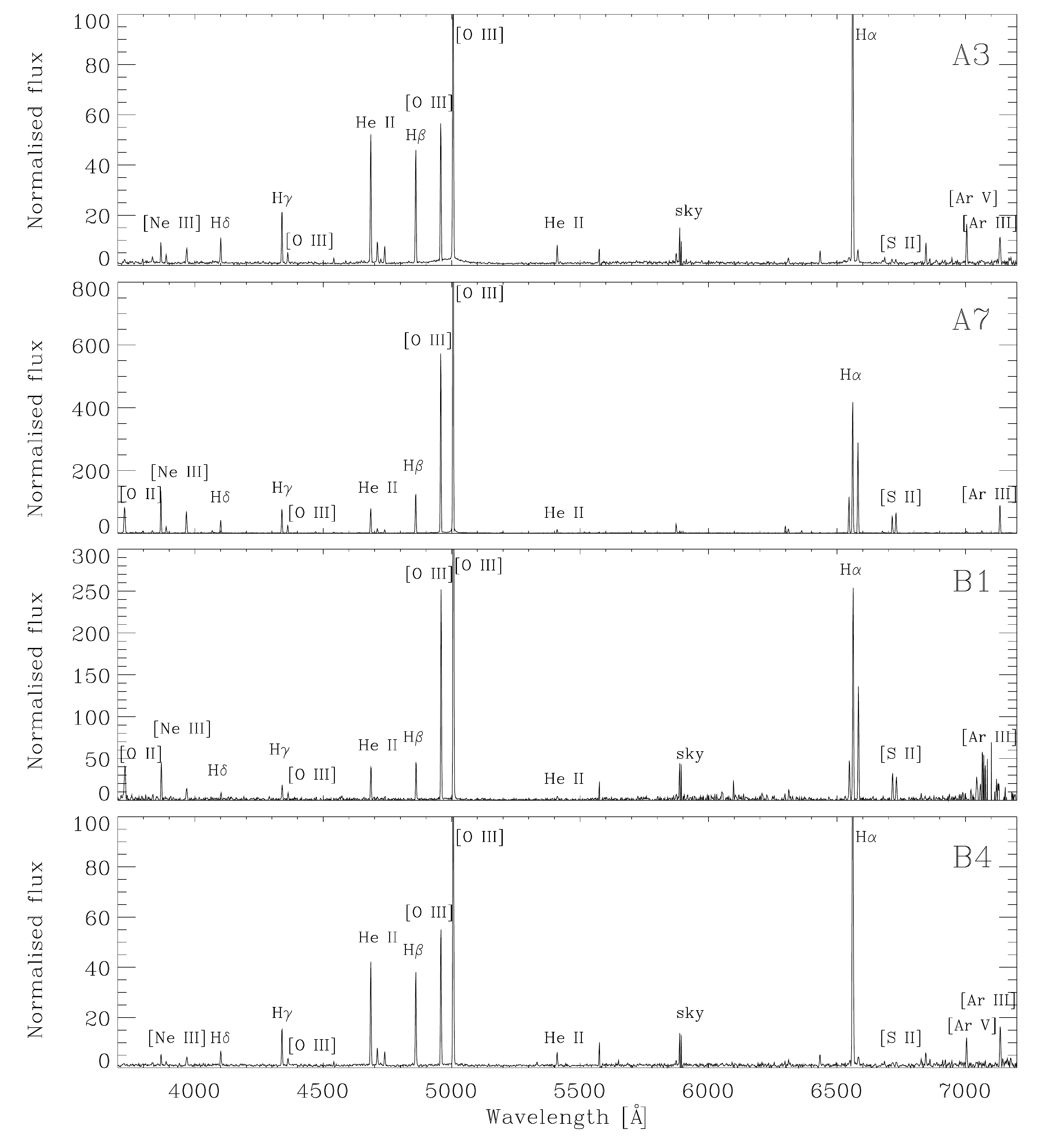}
\label{fig:spec_examples}
\caption{Examples of INT IDS spectra of NGC\,2371. The panels show the 
spectra extracted from regions A3, A7, B1 and B4. The most prominent lines 
are labeled. See Tables~ 3 and 4 for details on the fluxes.}
\end{center}
\end{figure*}

The parameters of our best-fit model are listed in Table~2.
We note that the stellar mass was adopted to be $M_{\star}=0.6$~M$_{\odot}$, a typical value for CSPNe \citep[e.g.,][and references therein]{MillerBertolami2016}. The resultant temperature, defined at $\tau_\text{Ross}=20$, which is constrained by the relative strength of the emission lines, is $T_\mathrm{eff}$=130~kK. A luminosity of $L_{\star}=2820$~L$_{\odot}$ is estimated by using a distance  $d=1.75$~kpc \citep{bailerjones2018} and the resultant mass-loss rate is $\dot{M}=1.78\times10^{-8}$~M$_{\odot}$~yr$^{-1}$.
We note that the estimated value for the mass-loss rate is somewhat smaller than those listed for other [WR] stars in \citet{Todt2013b}, who used a canonical luminosity of $L=5000\,L_\odot$ to calculate the mass-loss rates from the $R_\text{t}$ values obtained from the analysis. Taking the lower luminosity of NGC\,2371 into account, our value of $\dot{M}$ is comparable with the one derived for Hen\,2-55. Furthermore, our $R_\text{t}=20\,R_\odot$ is fully compatible with the value of $R_\text{t}=15^{+10}_{-5}\,R_\odot$ by \citet{Herald2004}. They used the same luminosity as in our analysis (based on the distance of $d=1.5\,$kpc), but assumed a smooth wind with $D=1$, where we adopted a density contrast of $D=10$ for a clumped wind. Therefore, their mass-loss rate is about a factor $\sqrt{10}$ larger than ours.

Our best-fit model to the P Cygni profiles in the UV spectra resulted in a stellar wind velocity $v_{\infty}=3700$~km~s$^{-1}$ with a micro-turbulence of less than 3\%. The micro-clumping parameter $D$, which is defined as the
density contrast between a smooth wind and a clumpy wind, is estimated to be 10. Figure~5 shows a comparison between the PoWR model of WD\,0722$+$295 and the optical and UV spectra, and Figure~6 presents the synthetic spectral energy distribution (SED) from the far-UV to the IR in comparison with UV and optical spectra and IR photometry obtained from public archives.

The abundances for WD\,0722$+$295 derived from our best-fit model are also listed in Table~2. Whereas the stellar parameters agree with those reported by \citet[][see their table~6]{Herald2004}, our abundance determinations are at variance, with our He 
abundance 30\% larger and our N/O ratio half their value.  
The differences can be attributed to the use of different stellar atmosphere codes, as \citet{Herald2004} use a relatively old version of the CMFGEN code \citep[][]{Hillier1998,Hillier1999}, but we note that our abundance determination is improved by the simultaneous modelling of high-quality optical spectra and UV data.

\section{Physical conditions, excitation and chemical abundances of NGC\,2371}

\begin{table*}
\centering
\caption[]{Reddeding-corrected line intensities for selected regions of Slit\,A of NGC\,2371}
\setlength{\columnwidth}{0.4\columnwidth}
\setlength{\tabcolsep}{0.7\tabcolsep}
\begin{tabular}{cccccccccc}
\hline
\multicolumn{1}{c}{$\lambda_{0}$}&
\multicolumn{1}{c}{line}&
\multicolumn{1}{c}{A1}&
\multicolumn{1}{c}{A2}&
\multicolumn{1}{c}{A3}&
\multicolumn{1}{c}{A5}&
\multicolumn{1}{c}{A6}&
\multicolumn{1}{c}{A7}&
\multicolumn{1}{c}{A8}&
\multicolumn{1}{c}{A9}\\
\multicolumn{1}{c}{}&
\multicolumn{1}{c}{}&
\multicolumn{1}{c}{(5$''$)}&
\multicolumn{1}{c}{(10$''$)}&
\multicolumn{1}{c}{(4$''$)}&
\multicolumn{1}{c}{(5$''$)}&
\multicolumn{1}{c}{(5$''$)}&
\multicolumn{1}{c}{(5$''$)}&
\multicolumn{1}{c}{(5$''$)}&
\multicolumn{1}{c}{(10$''$)}\\
\hline
3726.2   &  [O\,{\sc ii}]  & 33.2$\pm$5.4  & 35.3$\pm$1.8  & 11.6$\pm$2.4  & 16.5$\pm$3.0  & 24.1$\pm$1.7  & 201.2$\pm$4.9  & 141.7$\pm$8.1 & \dots \\
3797.9   &  H$\theta$      & 8.4$\pm$3.2   & 6.5$\pm$0.7   & 6.9$\pm$1.7   & 7.2$\pm$1.5   & 5.7$\pm$0.6   & 8.6$\pm$0.8    & \dots         & \dots \\
3835.4   &  H$\eta$        & 13.5$\pm$3.9  & 9.6$\pm$0.8   & 12.3$\pm$2.1  & 12.2$\pm$1.8  & 9.9$\pm$0.6   & 9.3$\pm$0.7    & \dots         & \dots \\
3868.8   &  [Ne\,{\sc iii}]& 99.4$\pm$7.0  & 81.3$\pm$1.9  & 32.6$\pm$2.2  & 33.5$\pm$2.2  & 74.7$\pm$1.2  & 157.4$\pm$2.3  & 97.5$\pm$6.1  & \dots \\
3889.1   &  H$\zeta$       & 16.5$\pm$3.5  & 16.0$\pm$0.9  & 12.6$\pm$1.7  & 14.9$\pm$1.5  & 16.2$\pm$0.6  & 21.9$\pm$0.8   & 15.6$\pm$2.7  & \dots \\
3967.5   &  [Ne\,{\sc iii}]& 49.8$\pm$5.0  & 44.3$\pm$1.4  & 24.4$\pm$2.2  & 29.5$\pm$2.0  & 43.7$\pm$1.0  & 72.1$\pm$1.4   & 46.1$\pm$3.7  & \dots \\
3970.0   &  H$\epsilon$    & \dots         & \dots         & \dots         & \dots         & \dots         & \dots          & \dots         & \dots \\
4026.2   &  He\,{\sc i}    & 4.6$\pm$2.6   & 2.0$\pm$0.4   & 1.2$\pm$0.8   & 2.7$\pm$1.1   & 2.2$\pm$0.4   & 2.6$\pm$0.3    & 1.1$\pm$0.9   & \dots \\
4068.6   &  [S\,{\sc ii}]  & 2.3$\pm$1.5   & 2.8$\pm$0.5   & 2.2$\pm$1.0  & \dots         & \dots         & \dots          & \dots         & \dots \\
4101.7   &  H$\delta$      & 31.9$\pm$3.4  & 32.7$\pm$1.0  & 27.2$\pm$1.8  & 31.9$\pm$1.5  & 34.1$\pm$0.8  & 33.3$\pm$0.7   & 26.5$\pm$2.4  & \dots \\
4200.0   &  He\,{\sc ii}   & \dots         & \dots         & \dots         & \dots         & 2.3$\pm$0.2   & 1.2$\pm$0.2    & \dots         & \dots \\
4340.5   &  H$\gamma$      & 54.0$\pm$3.5  & 57.2$\pm$1.0  & 49.9$\pm$1.8  & 54.9$\pm$1.6  & 56.6$\pm$1.0  & 57.4$\pm$0.9   & 46.2$\pm$3.3  & \dots \\
4363.2   &  [O\,{\sc iii}] & 19.6$\pm$2.2  & 17.1$\pm$0.6  & 10.0$\pm$1.0  & 9.7$\pm$0.8   & 17.0$\pm$0.6  & 18.3$\pm$0.4   & 16.0$\pm$1.8  & \dots \\
4471.5   &  He\,{\sc i}    & \dots         & \dots         & \dots         & \dots         & \dots         & 3.7$\pm$0.2    & 3.8$\pm$1.1   & \dots \\
4540.0   &  He\,{\sc ii}   & 3.8$\pm$1.1   & 4.3$\pm$0.4   & 3.4$\pm$0.8   & 4.1$\pm$0.7   & 4.4$\pm$0.5   & 2.4$\pm$0.3    & 2.4$\pm$0.8   & \dots \\
4686.0   &  He\,{\sc ii}   & 112.8$\pm$3.6 & 116.4$\pm$1.6 & 122.8$\pm$2.9 & 133.0$\pm$2.5 & 118.5$\pm$1.4 & 65.2$\pm$0.9   & 78.0$\pm$2.9  & \dots \\
4711.4   &  [Ar\,{\sc iv}] & 22.6$\pm$1.8  & 24.2$\pm$0.6  & 20.9$\pm$1.3  & 20.4$\pm$1.0  & 25.2$\pm$0.6  & 11.2$\pm$0.4   & 14.5$\pm$1.4  & \dots \\
4740.2   &  [Ar\,{\sc iv}] & 14.5$\pm$1.4  & 17.7$\pm$0.6  & 14.8$\pm$1.1  & 14.5$\pm$0.9  & 18.5$\pm$0.6  & 8.1$\pm$0.4    & 10.3$\pm$1.2  & \dots \\
4861.4   &  H$\beta$       & 100$\pm$2.2   & 100$\pm$0.9   & 100$\pm$1.8   & 100$\pm$1.6   & 100$\pm$0.7   & 100$\pm$0.5    & 100$\pm$2.6   & 100$\pm$   \\
4958.9   &  [O\,{\sc iii}] & 305.2$\pm$2.2 & 213.6$\pm$8.1 & 122.4$\pm$1.0 & 118.7$\pm$0.6 & 210.4$\pm$9.3 & 392.3$\pm$11.9 & 283.1$\pm$27.2& 453.1$\pm$152.8 \\
5006.8   &  [O\,{\sc iii}] & 907.0$\pm$5.7& 638.5$\pm$5.7 & 366.5$\pm$2.6 & 354.2$\pm$1.5 & 628.5$\pm$6.8 & 1174.7$\pm$8.6 & 849.4$\pm$27.9 & 1130.4$\pm$419.3\\
5197.9   &  [N\,{\sc i}]   & \dots         & \dots         & \dots         & \dots         & \dots         & 1.8$\pm$0.1    & 1.1$\pm$0.3   & \dots \\
5411.0   &  He\,{\sc ii}   & 6.2$\pm$0.7   & 8.1$\pm$0.2   & 7.4$\pm$0.5   & 8.1$\pm$0.4   & 8.3$\pm$0.2   & 4.6$\pm$0.1    & 4.6$\pm$0.6   & \dots \\
5517.7   &  [Cl\,{\sc iii}]& \dots         & 1.0$\pm$0.1   & \dots         & \dots         & 1.0$\pm$0.1   & 1.4$\pm$0.1    & \dots         & \dots \\
5537.9   &  [Cl\,{\sc iii}]& \dots         & 0.7$\pm$0.1   & \dots         & \dots         & 0.8$\pm$0.1   & 1.1$\pm$0.1    & \dots         & \dots \\
5754.6   &  [N\,{\sc ii}]  & \dots         & 0.4$\pm$0.1   & \dots         & \dots         & \dots         & 2.3$\pm$0.1    & 2.2$\pm$0.6   & \dots \\
5875.6   &  He\,{\sc i}    & 0.9$\pm$0.2   & 2.0$\pm$0.1   & 1.7$\pm$0.2   & 1.7$\pm$0.2   & 1.8$\pm$0.1   & 6.8$\pm$0.2    & 4.3$\pm$0.6   & \dots \\
6101.0   &  [K\,{\sc iv}]  & \dots         & 0.5$\pm$0.1   & \dots         & \dots         & 0.7$\pm$0.1   & 0.2$\pm$0.1    & 0.6$\pm$0.4   & \dots \\
6234.0   &  He\,{\sc ii}   & \dots         & \dots         & \dots         & \dots         & \dots         & \dots          & \dots         & \dots \\
6312.1   &  [S\,{\sc iii}] & 3.8$\pm$0.4   & \dots         & 2.1$\pm$0.4   & 1.9$\pm$0.3   & 4.2$\pm$0.1   & 5.6$\pm$0.1    & 3.9$\pm$0.7   & \dots \\
6363.0   &  [O\,{\sc i}]   & \dots         & 0.4$\pm$0.1   & \dots         & \dots         & \dots         & 3.3$\pm$0.1    & 1.8$\pm$0.5   & \dots \\
6406.0   &  He\,{\sc ii}   & \dots         & 1.1$\pm$0.2   & \dots         & 1.0$\pm$0.5   & 1.2$\pm$0.2   & 0.5$\pm$0.2    & \dots         & \dots \\
6435.1   &  [Ar\,{\sc v}]  & 2.1$\pm$0.6   & 3.8$\pm$0.3   & 5.6$\pm$0.7   & 6.7$\pm$0.6   & 4.0$\pm$0.2   & 0.9$\pm$0.1    & 1.9$\pm$0.9   & \dots \\
6548.1   &  [N\,{\sc ii}]  & 4.5$\pm$0.2  & 8.3$\pm$0.1  & 2.8$\pm$0.1  & 1.8$\pm$0.0  & 4.4$\pm$0.1  &  66.2$\pm$1.1   & 52.0$\pm$3.4  & \dots \\
6562.8   &  H$\alpha$      & 287.0$\pm$5.5 & 287.0$\pm$1.6 & 287.0$\pm$4.1 & 287.0$\pm$3.1 & 287.0$\pm$1.4 & 287.0$\pm$1.5  & 286.5$\pm$6.4 & 248.5$\pm$64.3 \\
6583.5   &  [N\,{\sc ii}]  & 15.8$\pm$1.3  &25.0$\pm$0.5  &7.9$\pm$0.7  &6.0$\pm$0.4  &15.3$\pm$0.3  & 191.6$\pm$1.2  &149.9$\pm$4.5 & \dots \\
6678.2   &  He\,{\sc i}    & \dots         & \dots         & \dots         & \dots         & \dots         & 4.5$\pm$0.3    & 2.4$\pm$0.9   & \dots \\
6716.5   &  [S\,{\sc ii}]  & 8.0$\pm$1.6   & 7.9$\pm$0.4   & 3.3$\pm$1.0   & 2.2$\pm$0.8   & 6.1$\pm$0.3   & 46.0$\pm$0.8   & 51.7$\pm$3.3  & \dots \\
6730.8   &  [S\,{\sc ii}]  & 6.6$\pm$1.5   & 8.2$\pm$0.4   & 2.5$\pm$0.9   & 1.8$\pm$0.9   & 5.9$\pm$0.3   & 58.2$\pm$0.9   & 50.4$\pm$3.2  & \dots \\
7005.9   &  [Ar\,{\sc v}]  & 15.6$\pm$4.0  & 25.6$\pm$1.3  & 41.9$\pm$4.0  & 43.6$\pm$3.2  & 29.0$\pm$1.0  & 7.5$\pm$0.5    & 15.4$\pm$4.5  & \dots \\
7065.2   &  He\,{\sc i}    & \dots         & 2.3$\pm$0.6   & \dots         & \dots         & 2.8$\pm$0.5   & 9.2$\pm$0.5    & \dots         & \dots \\
7135.8   &  [Ar\,{\sc iii}]& 78.7$\pm$6.6  & 84.2$\pm$1.7  & 40.0$\pm$3.7  & 30.0$\pm$2.4  & 75.8$\pm$1.3  & 158.2$\pm$1.7  & 99.0$\pm$9.1  & \dots \\
\hline
log(H$\beta$) &[erg~cm$^{-2}$~s$^{-1}$] & $\mathrm{-13.26}$& $\mathrm{-12.55}$ &$\mathrm{-13.45}$&$\mathrm{-13.14}$& $\mathrm{-12.68}$&$\mathrm{-12.73}$& $\mathrm{-13.65}$&$\mathrm{-15.24}$ \\ 
$c$(H$\beta$)                     &             & 0.38  & 0.16  & 0.06   & 0.22   &  0.17 & 0.08  & 0  & \dots \\
A$_{\rm V}$ & & 0.80 & 0.32 & 0.13 & 0.47 & 0.35& 0.16 & 0 & 0\\
\hline
$T_\mathrm{e}$([O\,{\sc iii}]) & [K] & 
16000$\pm$800 & 17800$\pm$200 & 18000$\pm$900 & 
18000$\pm$800 & 18000$\pm$200 & 13800$\pm$100 & 
15000$\pm$500 & \dots \\
$T_\mathrm{e}$([N\,{\sc ii}])   &   [K]     & \dots             & 11000$\pm$1100& \dots             & \dots             & \dots             & 9330$\pm$120&10300$\pm$1100& \dots \\
$n_\mathrm{e}$ ([S\,{\sc ii}]) & [cm$^{-3}$] & 
250$\pm$60 & 850$\pm$50 & 100$\pm$80 & 280$\pm$200 & 680$\pm$40 & 1640$\pm$40 & 620$\pm$50 & \dots \\
$n_\mathrm{e}$ ([Cl\,{\sc iii}])&[cm$^{-3}$]& \dots             & \dots              & \dots             & \dots             &270$\pm$220  & 880$\pm$80  & \dots              & \dots \\
$n_\mathrm{e}$ ([Ar\,{\sc iv}]) &[cm$^{-3}$]& \dots             &420$\pm$80    & \dots   &50$\pm$170   &440$\pm$90   & 330$\pm$130 &100$\pm$180    & \dots \\     
\hline
\end{tabular}
\vspace{0.4cm}
\label{sample}
\end{table*}

\begin{table*}
\centering
\caption[]{Reddeding-corrected line intensities for selected regions of Slit\,B of NGC\,2371}
\setlength{\columnwidth}{0.3\columnwidth}
\setlength{\tabcolsep}{0.4\tabcolsep}
\begin{tabular}{ccccccccccc}
\hline
\multicolumn{1}{c}{$\lambda_{0}$}&
\multicolumn{1}{c}{line}&
\multicolumn{1}{c}{B1}&
\multicolumn{1}{c}{B2}&
\multicolumn{1}{c}{B3}&
\multicolumn{1}{c}{B4}&
\multicolumn{1}{c}{B5}&
\multicolumn{1}{c}{B6}&
\multicolumn{1}{c}{B7}&
\multicolumn{1}{c}{B8}& 
\multicolumn{1}{c}{B9}\\
\multicolumn{1}{c}{}&
\multicolumn{1}{c}{}&
\multicolumn{1}{c}{(15$''$)}&
\multicolumn{1}{c}{(10$''$)}&
\multicolumn{1}{c}{(10$''$)}&
\multicolumn{1}{c}{(10$''$)}&
\multicolumn{1}{c}{(10$''$)}&
\multicolumn{1}{c}{(10$''$)}&
\multicolumn{1}{c}{(10$''$)}&
\multicolumn{1}{c}{(10$''$)}& 
\multicolumn{1}{c}{(15$''$)}\\
\hline
3726.2   & [O\,{\sc ii}]   & 258.7$\pm$41.7 & \dots           & \dots           & \dots         & \dots         & \dots         & \dots          & \dots           & 145.9$\pm$32.4 \\
3868.8   & [Ne\,{\sc iii}] & 190.1$\pm$35.2 & 100.6$\pm$37.2  & 100.1$\pm$29.5  & 36.8$\pm$3.8  & \dots         & 41.4$\pm$2.6  & \dots          & \dots           & 175.8$\pm$35.9 \\
3888.7   & He\,{\sc i}     & \dots          & \dots           & \dots           & 12.2$\pm$3.1  & \dots         & 14.1$\pm$2.0  & \dots          & \dots           & \dots          \\
3967.5   & [Ne\,{\sc iii}] & 73.9$\pm$18.9  & \dots           & \dots           & 29.8$\pm$3.7  & \dots         & 27.9$\pm$2.3  & \dots          & \dots           & 73.3$\pm$21.6  \\
4101.7   &  H$\delta$      & 31.4$\pm$11.6  & \dots           & \dots           & 30.7$\pm$3.0  & 26.6$\pm$2.4  & 34.2$\pm$2.1  & 38.4$\pm$13.7  & \dots           & 30.2$\pm$11.6   \\
4340.5   &  H$\gamma$      & 53.0$\pm$10.8  & 57.9$\pm$18.7   & 36.9$\pm$12.3   & 56.5$\pm$3.2  & 90.0$\pm$2.9  & 58.5$\pm$2.0  & 51.5$\pm$14.3  & 72.5$\pm$23.4   & 55.7$\pm$10.7   \\
4363.2   &  [O\,{\sc iii}] & 21.6$\pm$7.7   & 44.5$\pm$17.1   & \dots           & 11.7$\pm$2.0  & 10.6$\pm$3.1  & 12.0$\pm$1.2  & \dots          & \dots           & 14.4$\pm$6.5   \\
4540.0   &  He\,{\sc ii}   & \dots          & \dots           & \dots           & \dots         & \dots         & 3.8$\pm$0.8   & \dots          & \dots           & \dots          \\
4686.0   &  He\,{\sc ii}   & 96.9$\pm$13.3  & 107.6$\pm$19.9  & 107.6$\pm$15.4  & 130.8$\pm$4.0 & \dots         & 133.1$\pm$3.2 & 126.1$\pm$12.0 & 138.4$\pm$22.7  & 103.3$\pm$12.9  \\
4711.4   &  [Ar\,{\sc iv}] & \dots          & \dots           & 15.4$\pm$8.6    & 22.4$\pm$2.0  & \dots         & 20.9$\pm$1.1  & \dots          & \dots           & \dots          \\
4740.2   &  [Ar\,{\sc iv}] & \dots          & \dots           & 12.6$\pm$7.9    & 16.1$\pm$1.7  & 14.6$\pm$1.6  & 17.3$\pm$0.9  & \dots          & \dots           & \dots          \\
4861.4   &  H$\beta$       & 100$\pm$9.6    & 100$\pm$16.9    & 100$\pm$12.2    & 100$\pm$3.4   & 100$\pm$1.3   & 100$\pm$1.8   & 100$\pm$8.8  & 100$\pm$15.6& 100$\pm$10.6       \\
4958.9   &  [O\,{\sc iii}] & 533.3$\pm$21.2 & 407.2$\pm$28.1  & 278.2$\pm$17.7  & 135.7$\pm$3.8 & 154.7$\pm$1.9 & 135.1$\pm$2.1 & 253.5$\pm$12.5 & 329.5$\pm$23.5  & 478.5$\pm$22.3 \\
5006.8   &  [O\,{\sc iii}] & 1549.7$\pm$31.1& 1191.3$\pm$45.1 & 807.0$\pm$26.0  & 401.9$\pm$6.0 & 448.3$\pm$3.1 & 394.9$\pm$3.4 & 711.3$\pm$21.5 & 954.6$\pm$41.3  & 1385.8$\pm$38.6\\
5411.0   &  He\,{\sc ii}   & 5.7$\pm$2.2    & \dots           & \dots           & 6.5$\pm$0.6   & 8.1$\pm$0.9   & 7.5$\pm$0.4   & 7.4$\pm$2.5    & \dots           & 5.3$\pm$2.1    \\
5875.6   &  He\,{\sc i}    & \dots          & \dots           & \dots           & 2.0$\pm$0.4   & \dots         & 2.1$\pm$0.2   & \dots          & \dots           & \dots          \\
6101.0   &  [K\,{\sc iv}]  & \dots          & \dots           & \dots           & \dots         & \dots         & \dots         & \dots          & \dots           & \dots          \\
6312.1   &  [S\,{\sc iii}] & 5.8$\pm$2.9    & \dots           & \dots           & 1.7$\pm$0.6   & \dots         & 2.0$\pm$0.3   & \dots          & \dots           & 4.8$\pm$1.9    \\
6435.1   &  [Ar\,{\sc v}]  & \dots          & \dots           & \dots           & 4.3$\pm$0.9   & 3.8$\pm$1.7   & 4.2$\pm$0.4   & \dots          & \dots           & \dots          \\
6548.1   &  [N\,{\sc ii}]  & 46.0$\pm$6.3& 24.5$\pm$11.1& 18.2$\pm$2.8 & 3.4$\pm$0.2 & \dots & 1.0$\pm$0.1& \dots & \dots        & 31.8$\pm$5.1   \\
6562.8   &  H$\alpha$      & 287.0$\pm$17.7 & 287.0$\pm$26.3  & 287.0$\pm$17.4  & 287.0$\pm$6.8 & 287.0$\pm$3.9 & 287.0$\pm$4.3 & 287.0$\pm$17.6 & 287.0$\pm$32.2  & 287.0$\pm$15.0 \\
6583.5   &  [N\,{\sc ii}]  & 159.0$\pm$13.0 & 49.3$\pm$11.7 &43.0$\pm$7.0 & 6.2$\pm$1.3   & 12.5$\pm$2.7  & 3.7$\pm$0.5   & 9.4$\pm$4.1    & 21.0$\pm$10.0   & 106.6$\pm$9.0  \\
6716.5   &  [S\,{\sc ii}]  & 46.6$\pm$9.9   & \dots           & 17.4$\pm$7.9    & 2.8$\pm$1.6   & 2.5$\pm$1.8   & 2.0$\pm$0.6   & \dots          & \dots           & 33.6$\pm$6.6   \\
6730.8   &  [S\,{\sc ii}]  & 41.4$\pm$9.4   & \dots           & 18.0$\pm$8.2    & 2.8$\pm$1.5   & 2.9$\pm$2.3   & 2.2$\pm$0.6   & \dots          & \dots           & 31.1$\pm$6.5   \\
7005.9   &  [Ar\,{\sc v}]  & \dots          & \dots           & \dots           & 28.0$\pm$5.1  & 23.1$\pm$4.8  & 27.8$\pm$2.8  & \dots          & \dots           & \dots          \\
7135.8   &  [Ar\,{\sc iii}]& 115.7$\pm$2.1  & 127.3$\pm$55.3  & 62.5$\pm$28.4   & 49.5$\pm$5.9  & 37.3$\pm$5.9  & 28.1$\pm$2.6  & 50.0$\pm$17.8  & \dots           & 98.9$\pm$22.5  \\
\hline
log(H$\beta$)       &[erg~cm$^{-2}$~s$^{-1}$] & $\mathrm{-13.38}$&$\mathrm{-14.10}$& $\mathrm{-13.87}$&$\mathrm{-13.13}$& $\mathrm{-12.88}$&$\mathrm{-12.66}$&$\mathrm{-13.67}$& $\mathrm{-14.14}$& $\mathrm{-13.43}$\\
$c$(H$\beta$)                     &             & 0.49  & 0.41  & 0.46  & 0.35  & 0.56  & 0.54  & 0.54  & 0.44  & 0.47  \\
A$_{\rm V}$ & & 1.02& 0.86& 0.98& 0.74& 1.17& 1.14& 1.13& 0.91 & 0.99\\
\hline
$T_\mathrm{e}$([O\,{\sc iii}]) & [K] & 
13100$\pm$1700 & 22000$\pm$5000 & \dots & 18600$\pm$1600 & 15100$\pm$2300 & 19000$\pm$1000 & \dots & \dots & 11800$\pm$1800 \\ 
$n_\mathrm{e}$([S\,{\sc ii}]) & [cm$^{-3}$]& 380$\pm$30 & \dots          &690: & 810$\pm$360 & 1190$\pm$240 & 1070$\pm$30 & \dots & \dots &450$\pm$30  \\
$n_\mathrm{e}$ ([Ar\,{\sc iv}])&[cm$^{-3}$]& \dots              & \dots          & 1600$\pm$500 & 140$\pm$200   & \dots              & 2160$\pm$40 & \dots & \dots & \dots     \\
\hline
\end{tabular}
\vspace{0.4cm}
\label{sample}
\end{table*} 

Examples of INT IDS spectra extracted from Slits A and B are shown in Figure~7.
The complete list of lines detected in the different extraction regions 
and their dereddened intensities relative to an arbitrary value 100 for 
H$\beta$ are presented in Tables~3 and 4. 
The spectra extracted from Slit~B, particularly those from regions B2, B3, 
B7, and B8, show fewer emission lines than those extracted from Slit~A, in 
agreement with the lower surface brightness of the major axis probed by 
Slit~B (see Fig.~1). 
Intensity uncertainties at 1-$\sigma$ are provided in these tables, together 
with the values derived for $c$(H$\beta$) and the corresponding fluxes of 
the H$\beta$ line for each extraction region. 

Different line intensity ratios have been used to determine the physical conditions of the different regions in NGC\,2371 using the {\sc iraf} task {\it temden} \citep{Tody1993}.
$T_\mathrm{e}$ was estimated using the [O\,{\sc iii}] emission lines for 
most regions and the [N\,{\sc ii}] emission lines whenever possible, i.e., 
in regions A2, A7, and A8.  
$n_\mathrm{e}$ was derived using the 
[S\,{\sc ii}] $\lambda\lambda$6717,6731, 
[Cl\,{\sc iii}] $\lambda\lambda$5517,5537, and 
[Ar\,{\sc iv}] $\lambda\lambda$4711,4740 doublets whenever available. 
The values of $T_\mathrm{e}$ and $n_\mathrm{e}$, together with their 1-$\sigma$
uncertainties are listed in the bottom rows of Tables~3 and 4.

\begin{figure}
\begin{center}
\includegraphics[angle=0,width=1.0\linewidth]{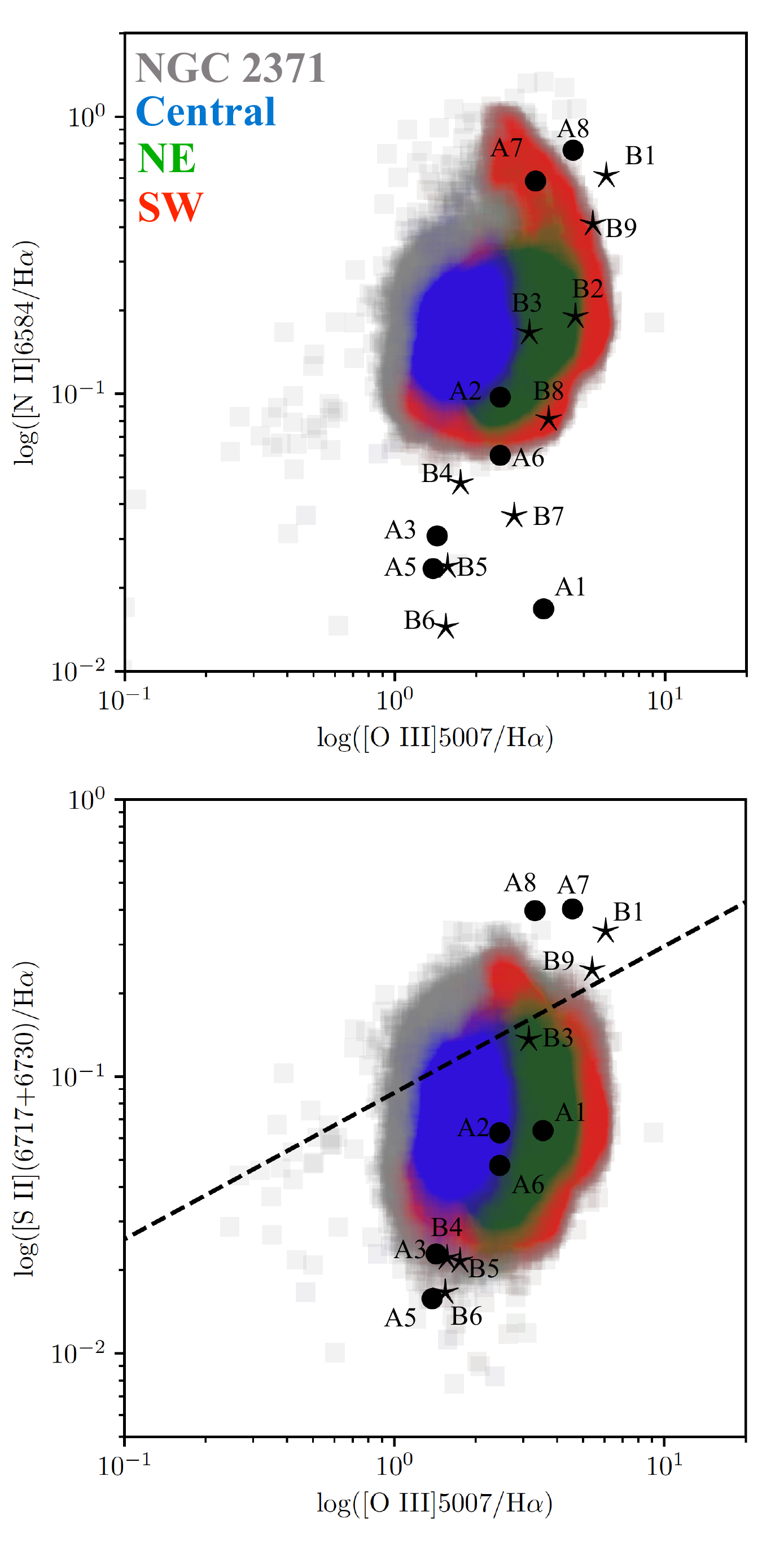}
\label{fig:ionisation}
\caption{Line intensity ratios derived for spectra extracted from the regions in Slit~A (bullets) and Slit~B (stars), and from different regions of \emph{HST} flux-calibrated ratio maps as described by the coloured points. Different colours represent line ratios extracted from the regions defined in Figure~9 left panel. 
The dashed line in the bottom panel marks the theoretical limit 
  $\log$([O\,{\sc iii}]/H$\alpha$) = 1.89 $\times\;\log$([S\,{\sc ii}]/H$\alpha$)\;+\;2.46 
  (see Section~4 for details) between photoionization (below the line) and shock-excitation
  (above the line).}
\end{center}
\end{figure}
 
  The values of $T_\mathrm{e}$([O\,{\sc iii}]) listed in Tables~3 and 4 
  reveal notable temperature gradients in the central cavity of NGC\,2371.  
  The innermost regions A2, A3, A5, A5, B4, and B6 immediately around the 
  CSPN show consistently high values of $T_\mathrm{e}$([O\,{\sc iii}]) in 
  the range 18,000--19,000~K,
  that decrease to $\approx$15,000--16,000~K
  in the outermost regions A1 and A8, 
  and have the notably lowest value of $\approx$13,800~K
  in the region A7, 
  which probes the SW knot.
  
Such high values of $T_\mathrm{e}$ in the close vicinity of the CSPN of 
NGC\,2371 arise from its 130~kK high $T_\mathrm{eff}$ (see Section~3.1). 
  Interestingly, $T_\mathrm{e}$([N\,{\sc ii}]) is only computed on  
  regions that probe the NE and SW knots.  
  The much lower values of $T_\mathrm{e}$([N\,{\sc ii}]) in these regions 
  with respect to those of $T_\mathrm{e}$([O\,{\sc iii}]) indicate that 
  these apertures probe a mix of low- and high-excitation material.

  As for $n_\mathrm{e}$([S\,{\sc ii}]), it varies in Tables~3 and 4 from
  100~cm$^{-3}$ to 1640~cm$^{-3}$,
  with the densest region corresponding 
  to the core of the SW knot A7, whose head A6 and tail A8 have intermediate density values 600--700~cm$^{-3}$.
The NE knot seems to have lower density estimates than the SW know, but we note that Slit~A does not goes exactly across the NE knot. 
Furthermore, the NE knot is smaller and appears more fragmented than the 
SW knot (see Fig.~1 right panels). 
These facts will have consequences for the abundance determination too 
(see below). 

Whereas the regions A3, A5, B4 and B6 immediately around the CSPN share 
similar values of $T_{\rm e}$, the densities $n_{\rm e}$([S\,{\sc ii}]) 
are larger for regions B4 and B6 ($n_\mathrm{e}$= 810--1100~cm$^{-3}$) 
than for regions A3 and A5 (100--280~cm$^{-3}$).  
The latter seem to probe a hollow region around the CSPN. 
Regions B1 and B9 extracted from the outer edges of the major 
axis of NGC\,2371 have the lowest temperature values 
($T_\mathrm{e}$([O\,{\sc iii}])$\simeq$11,800--13,100~K), but 
they have higher densities 
($n_\mathrm{e}$([S\,{\sc ii}])$\approx400$~cm$^{-3}$) than 
regions within the inner main cavity.

In order to unveil differences in excitation from different morphological features in NGC\,2371, we plot in Figure~8 line ratios of key emission lines derived from the INT IDS spectroscopy \citep[see][and references therein]{Akras2016}, including 
[N\,{\sc ii}]~6583/H$\alpha$ vs.\ [O\,{\sc iii}]~5007/H$\alpha$ (top panel) 
and [S\,{\sc ii}]~(6717$+$6731)/H$\alpha$ vs.\ [O\,{\sc iii}]~5007/H$\alpha$ (bottom panel). 
Differences in excitation are clear: regions close to the star (inner regions -- A3, A5, B4, B5, B6) show the lowest [O\,{\sc iii}]/H$\alpha$ and [N\,{\sc ii}]/H$\alpha$ line ratios, whereas other regions, particularly the outermost regions B1 and B9, and the regions A7 and A8 of the SW knot show the highest [O\,{\sc iii}]/H$\alpha$ and [N\,{\sc ii}]/H$\alpha$.  

In Figure~8 bottom panel we show the theoretical limit between photoionised and shocked material $\log$([O\,{\sc iii}]/H$\alpha$) = 1.89~$\log$([S\,{\sc ii}]/H$\alpha$) + 2.46, as estimated by \citet{Danehkar2018} based on the models presented by \citet{Raga2008}. Line ratios below this limit show gas excited by ionisation, while regions above present shocked excited gas. Most morphological features in NGC\,2371 are located below this theoretical limit, i.e., they are excited by photoionisation.  
On the other hand, the outermost B1 and B9 regions are located above the dashed line, suggesting that they are excited by shocks as could be expected if these structures were expanding at high speed into the interstellar medium. 
The spectra of regions A7 and A8 extracted from the head of the SW dense knot also present line intensity ratios typical of shock-excitation. 

\begin{figure*}
\begin{center}
  \includegraphics[angle=0,width=\linewidth]{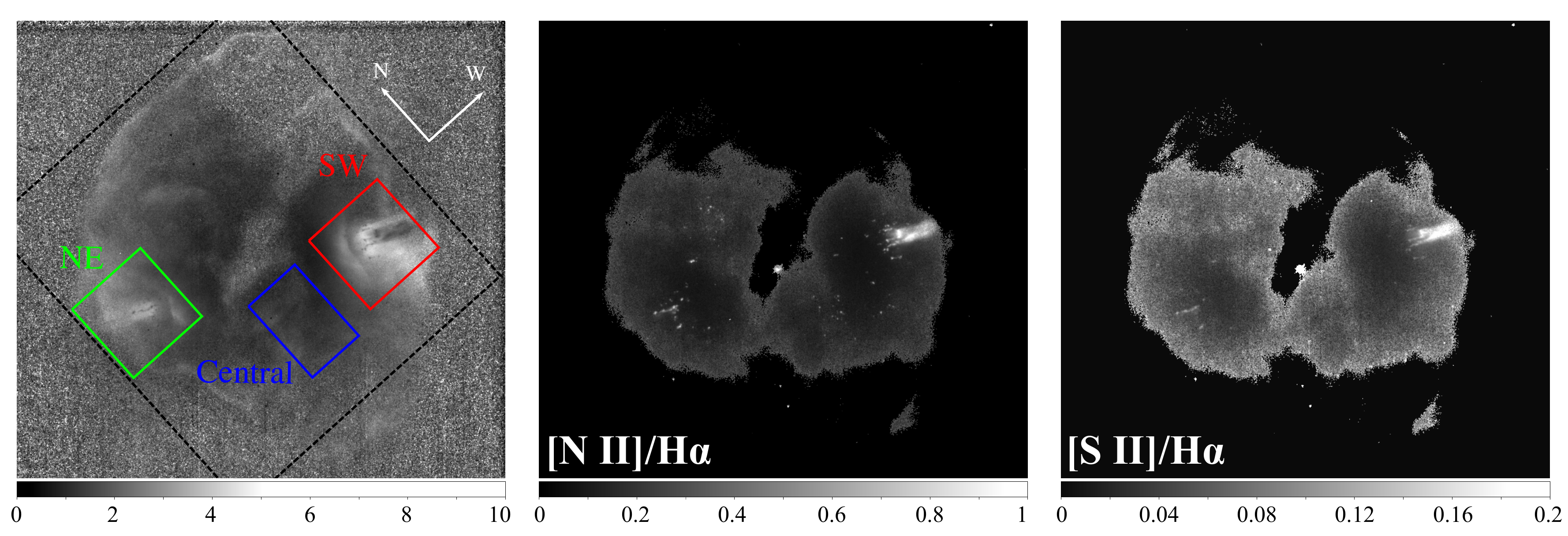}
\label{fig:ionisation}
\caption{Line intensity ratio maps obtained from calibrated {\it HST} images. ({\it Left}) The (black) dashed-line, red line, blue line, and green line rectangles represent different regions used to study the ionisation structure of NGC\,2371 in Figure~8. The three panels have the same field of view and orientation.}
\end{center}
\end{figure*}

The \emph{HST} WFC3 images of NGC\,2371 have also been used to provide a complementary 
view of its ionisation structure.
Ratio maps were obtained by dividing the [O\,{\sc iii}], [N\,{\sc ii}] and [S\,{\sc ii}] calibrated images by the H$\alpha$ image (see Fig.~9).  
We note that the \emph{HST} [S\,{\sc ii}] image does not have adequate signal-to-noise ratio in the NW and SE lobes of NGC\,2371. Thus, the line ratio calculations were only performed for the innermost regions of NGC\,2371. 
We defined four different rectangular regions (see Fig.~9 left panel) to study the dominant excitation mechanisms: 
a large region that includes most of the main inner cavity of NGC\,2371, 
two regions covering the SW and NE dense knots, and 
a region that includes only gas around the central star. 
We computed the pixel-by-pixel line ratios and these are plotted in Figure~8 as colour-coded regions alongside the results found for regions extracted from Slit~A and B. The results are consistent with  those obtained from the spectroscopic observations. 

Finally, the abundances were computed using the extensively tested code 
{\sc PyNEB} developed by \citet[][]{Luridiana2015}.  
PyNEB computes the physical conditions ($T_\mathrm{e}$ and $n_\mathrm{e}$) 
and ionic and total abundances.  
Ionic abundances were calculated with their corresponding temperature 
and averaged density based on the ionization potential (IP) of the ion. 
We note that the $T_\mathrm{e}$ and $n_\mathrm{e}$ values obtained from PyNEB 
(not presented here) are consistent within $<$2\% with those listed in Tables~3 
and 4 obtained in {\sc iraf}.
For the calculation of ionic abundances, we only considered emission 
lines with uncertainties smaller than 50\%, resulting in abundances 
with typical uncertainties of a few percent. 
The total abundances of He, O, N, Ne, Ar, S and Cl, listed in Table~5 for 
apertures A1--A3 and A5--A8, were computed adopting the ionization correction 
factors (ICFs) provided by \citet{Delgado2014}. 
The error budget of the abundances of heavy elements in 
Table~5 is mostly dominated by uncertainties in the 
computation of those ICFs, which are notably large 
for high excitation nebulae. 

\section{Discussion}

NGC\,2371 has a multi-component morphology with 
(1) a pair of outer lobes aligned along PA$=128^{\circ}$, 
(2) a high-excitation main shell with apparent elliptical morphology aligned along the same direction with two lower excitation caps at its tips, and 
(3) two ensembles of dense low-ionisation knots
in the equatorial region of the main shell, although 
misaligned with the previous structures. 
We provide next a description of the overall physical structure of NGC\,2371 using kinematical information to help us interpret its different components.  

\subsection{On the kinematic structure of NGC\,2371}

By Gaussian fitting of bright emission lines, their centroid can be routinely 
determined with an accuracy $\sim$10 times better than the spectral resolution, 
which is $\sim$200 km~s$^{-1}$ for our observations with the R400V grating and 
$\sim$70 km~s$^{-1}$ for those with the R1200U grating.  
Thus, our low resolution spectra allow us to investigate the kinematics of 
structures that have velocities $\geq$20 km~s$^{-1}$.
This is illustrated in Figure~10, which presents three sections of the  R400V 2D-spectrum obtained with the Slit B along the major nebular axis.
These spectral sections include the H$\alpha+$[N~{\sc ii}], [O~{\sc ii}] and [O~{\sc iii}] emission lines. Velocity gradients are clearly seen in these (as well as other) spectral lines as an S-shaped pattern in the position-velocity (PV) map along the nebular axis, but tilted along the opposite direction for the main cavity.  

These PV maps have been used to estimate the systemic radial velocities of the different morphological features in NGC\,2371 (i.e., the difference of the radial velocities of a feature with respect to the average radial velocity of the nebula) labeled in Figure~2 next to their corresponding extraction regions. 
The outermost structures B1 and B9 have systemic radial velocities of $\pm$30 km~s$^{-1}$, but the tips of the inner cavity B3 and B7 have larger systemic radial velocities of $\pm$70~km~s$^{-1}$. 
Meanwhile, regions B4 and B6 inside the main nebular shell have systemic radial velocities of $\pm$40 km~s$^{-1}$ with opposite sign than those of the adjacent tips B3 and B7. 
  The fast expansion of the bipolar lobes in NGC\,2371 is consistent with 
  the faster expansion of PNe with [WR]-type CSPNe compared to other PNe 
  \citep{Pena2003}.

\begin{figure}
\begin{center}
  \includegraphics[angle=0,width=\linewidth]{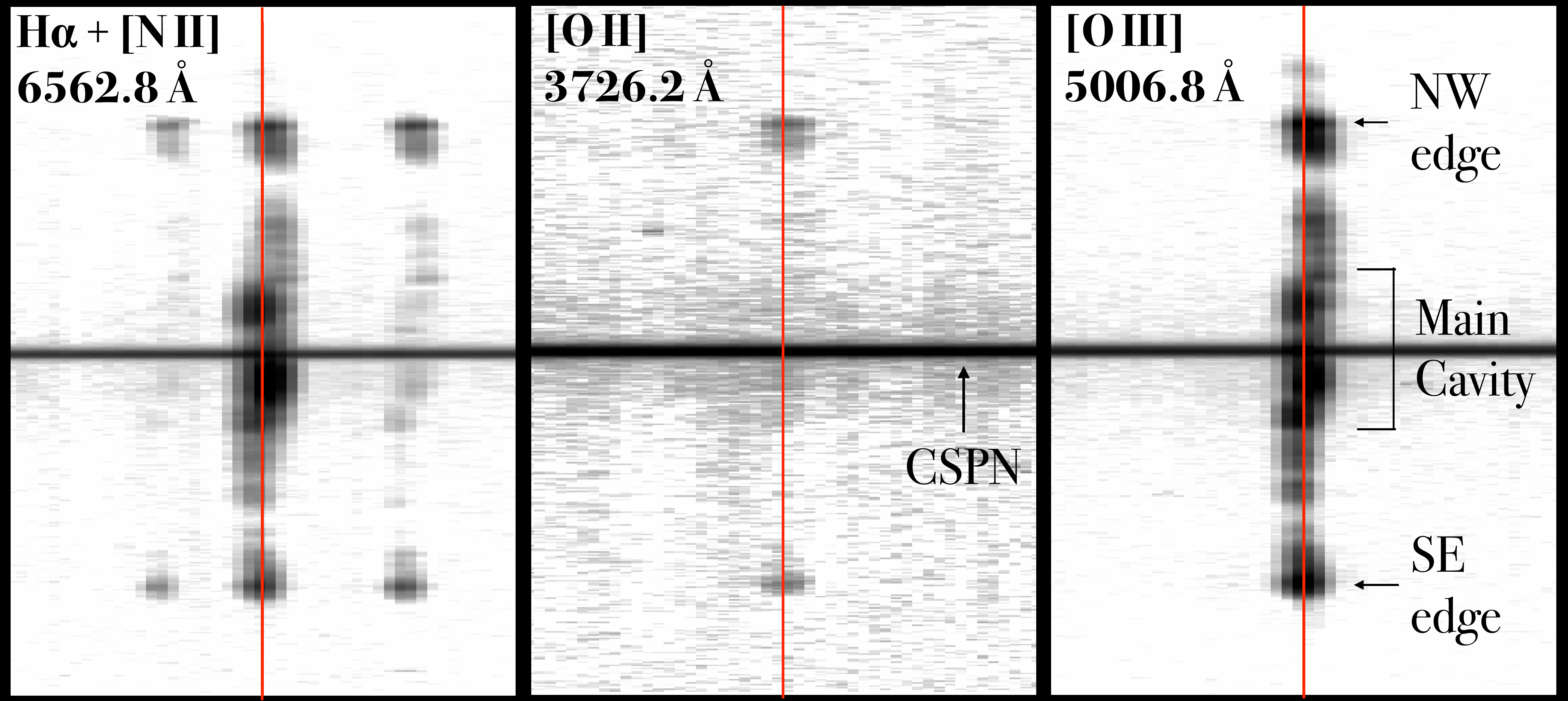}
\label{fig:lines_vel}
\caption{Sections of the INT IDS spectrum of the major axis of NGC\,2371 (Slit~B) centered on the H$\alpha+$[N\,{\sc ii}], [O\,{\sc ii}] and [O\,{\sc iii}] lines. The red vertical lines show the position of the rest wavelength of each line. The horizontal axis represents wavelength while the vertical the spatial profile.}
\end{center}
\end{figure}

\begin{figure}
\begin{center}
  \includegraphics[angle=0,width=1.0\linewidth]{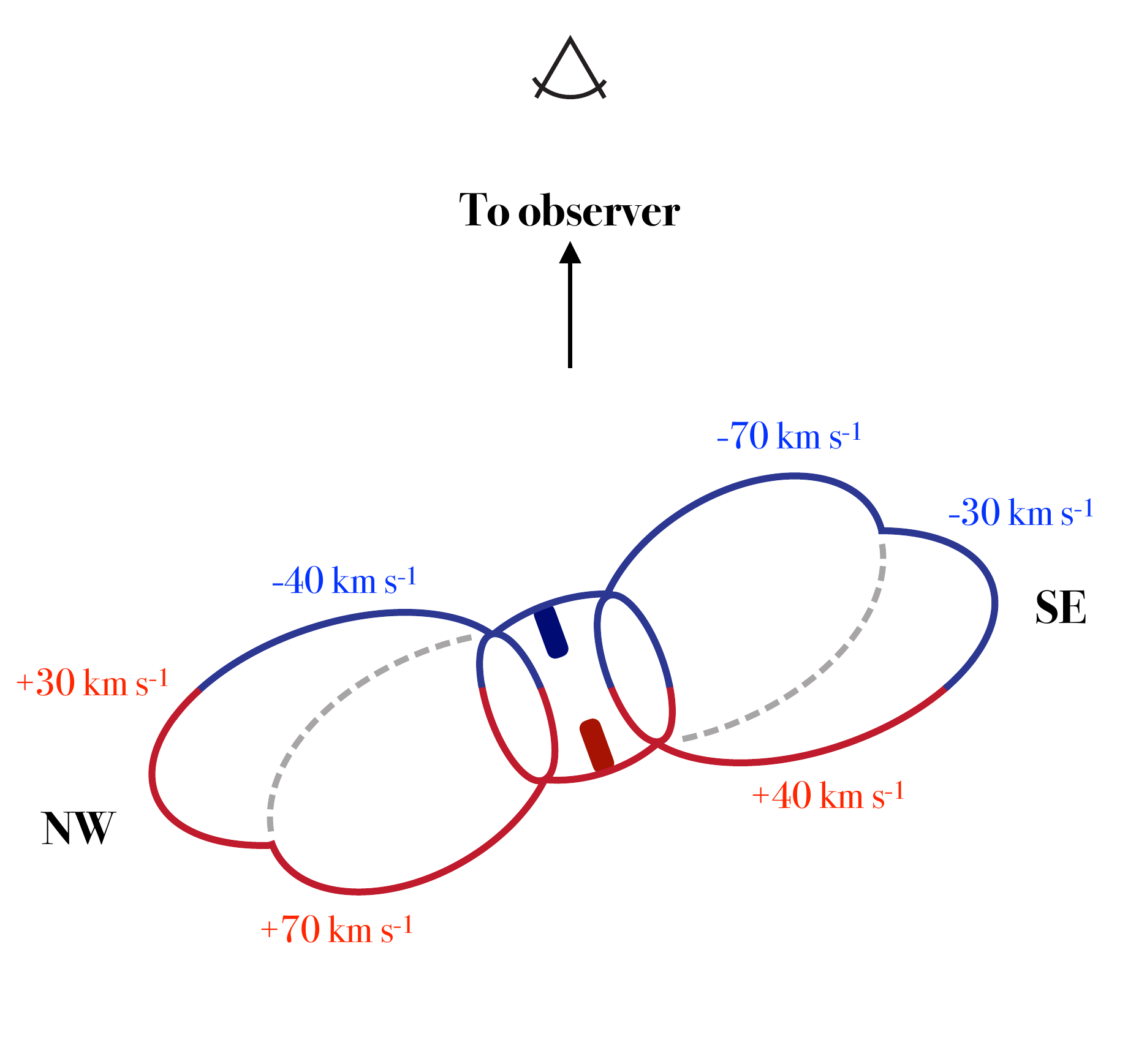}
\label{fig:vel_stru}
\caption{Velocity structure of NGC\,2371 suggested by the radial velocities obtained from our spectral analysis.}
\end{center}
\end{figure}

The overall expansion patterns described above are consistent with those reported by \citet{Ayala2005} using longslit high-dispersion echelle data. With that information at hand, we can envisage the simple model of the physical structure along the main nebular axis of NGC\,2371 sketched in Figure~11. The physical structure of NGC\,2371 shares many similarities with that of NGC\,650-1 \citep[see][]{RamosLarios2018}, with two pairs of bipolar lobes tilted with the line of sight along different inclination angles, 
and a toroidal or barrel-like central cavity.
In this geometrical model, the reversal of systemic radial velocities between 
the toroidal structure and the tips of the innermost bipolar lobes B3 and B7 
can be explained if the toroidal structure is orthogonal to the bipolar lobes.  

\begin{figure*}
\begin{center}
  \includegraphics[angle=0,width=1.0\linewidth]{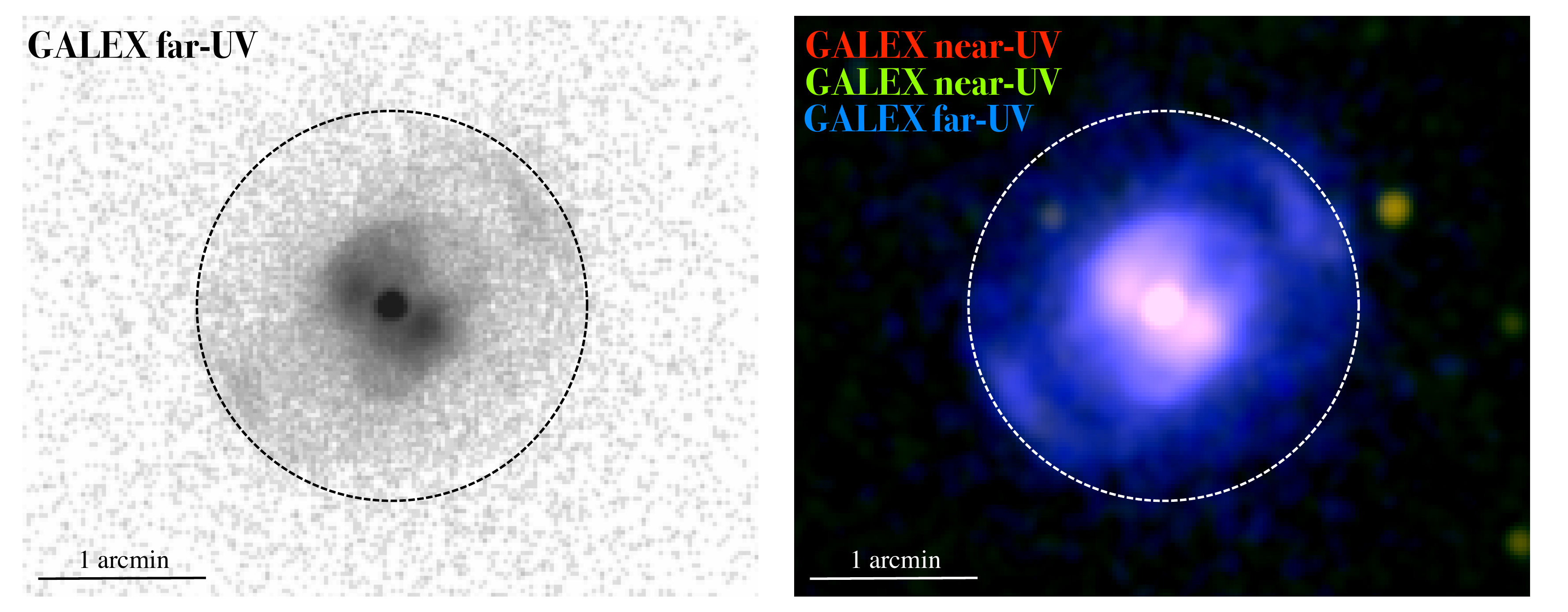}
\label{fig:compare}
\caption{{\it GALEX} UV images of NGC\,2371. ({\it Left}) {\it GALEX} far-UV gray scale image with the natural {\it GALEX} pixel size of 1.5~arcsec. ({\it Right}) Colour-composite UV image using the near- and far-UV filters. The dashed-line circle in both panels has an angular radius of 70~arcsec. North is up and east to the left.}
\end{center}
\end{figure*}

The present data allow a basic description of NGC\,2371, but we note that the NW and SE lobes have a rich morphology that suggests a more complex structure than that proposed in Figure~11.  
A complete analysis of the physical structure of NGC\,2371 using information obtained along different slit positions using the Manchester Echelle Spectrograph (MES) at the 2.1m telescope of the Observatorio Astron\'omico Nacional de San Pedro M\'artir (SPM-OAN) in conjunction with a morpho-kinematic model is underway (V\'{a}zquez et al., in preparation) and will certainly produce a detailed view of the physical structure of this PN to peer further into its formation history. Yet, the available data still have an interesting piece of information. 
The NE and SW low-ionisation knots move at systemic radial velocities 
$\pm7$~km~s$^{-1}$ considerably smaller than that of the toroidal 
structure of the main nebula where they seem to be embedded. 
These velocities, which are consistent with those reported by \citet{Sabbadin1982}, imply that the knots are being overtaken by the nebular material as suggested by \citet{RamosLarios2012}. Furthermore, the arc-like features observed around these structures (see Fig.~1 right panels) seems to suggest that they are interacting with the current fast wind from the CSPN, similarly to what is observed in {\it HST} images of Orion proplyds \citep[see][]{GA2001}. 

\begin{table*}
\centering
\caption[]{Total abundances for different regions in NGC\,2371}
\setlength{\columnwidth}{0.2\columnwidth}
\setlength{\tabcolsep}{0.55\tabcolsep}
\begin{tabular}{ccccccccc}
\hline
\multicolumn{1}{l}{} &
\multicolumn{1}{c}{A1}&
\multicolumn{1}{c}{A2}&
\multicolumn{1}{c}{A3}&
\multicolumn{1}{c}{A5}&
\multicolumn{1}{c}{A6}&
\multicolumn{1}{c}{A7}&
\multicolumn{1}{c}{A8}&
\multicolumn{1}{c}{Solar}\\
\multicolumn{1}{l}{} &
\multicolumn{1}{c}{}&
\multicolumn{1}{c}{}&
\multicolumn{1}{c}{}&
\multicolumn{1}{c}{}&
\multicolumn{1}{c}{}&
\multicolumn{1}{c}{}&
\multicolumn{1}{c}{}&
\multicolumn{1}{c}{\citep{Lodders2010}}\\
\hline
He  & 0.106$\pm$0.013 & 0.116$\pm$0.005 & 0.122$\pm$0.005 & 0.130$\pm$0.006 & 0.118$\pm$0.003 & 0.102$\pm$0.002 & 0.098$\pm$0.007 & 0.084 \\
O   & (7.1$\pm$3.0)$\times10^{-4}$ & (3.8$\pm$1.3)$\times10^{-4}$ & (1.6$\pm$0.6)$\times10^{-4}$ & (1.7$\pm$0.7)$\times10^{-4}$ & (2.8$\pm$0.9)$\times10^{-4}$ & (6.8$\pm$2.4)$\times10^{-4}$ & (4.9$\pm$1.7)$\times10^{-4}$ & 5.37$\times10^{-4}$ \\
N   & (2.2$\pm$1.4)$\times10^{-4}$ & (1.1$\pm$0.4)$\times10^{-4}$ & (9.2$\pm$3.2)$\times10^{-5}$ & (4.7$\pm$1.7)$\times10^{-5}$ & (1.3$\pm$0.4)$\times10^{-4}$ & (2.2$\pm$0.8)$\times10^{-4}$ & (2.1$\pm$0.7)$\times10^{-4}$ & 7.24$\times10^{-5}$ \\
Ne  & (3.4$\pm$1.4)$\times10^{-4}$ & (3.4$\pm$1.0)$\times10^{-4}$ & (8.1$\pm$2.4)$\times10^{-5}$ & (1.1$\pm$0.4)$\times10^{-4}$ & (1.6$\pm$0.5)$\times10^{-4}$ & (1.4$\pm$0.4)$\times10^{-5}$ & (1.8$\pm$0.5)$\times10^{-4}$ & 1.12$\times10^{-4}$ \\
Ar  &   3.5$\times10^{-5}$: & 1.5$\times10^{-5}$: & 1.1$\times10^{-5}$ & 5.0$\times10^{-6}$: & 1.2$\times10^{-5}$: & 1.3$\times10^{-5}$: & 9.3$\times10^{-6}$: & 3.16$\times10^{-6}$ \\
S   & (2.9$\pm$1.5)$\times10^{-5}$ & (7.2$\pm$2.9)$\times10^{-6}$& (8.5$\pm$3.4)$\times10^{-6}$ & (4.4$\pm$1.8)$\times10^{-6}$ & (1.5$\pm$0.6)$\times10^{-5}$ & (1.7$\pm$0.7)$\times10^{-5}$ & (1.7$\pm$0.7)$\times10^{-5}$ & 1.45$\times10^{-5}$ \\
Cl  & \dots & (1.9$\pm$0.7)$\times10^{-7}$ & \dots & \dots & (3.6$\pm$1.4)$\times10^{-7}$ & (1.5$\pm$0.6)$\times10^{-7}$ & \dots               & 1.78$\times10^{-7}$\\
\hline
He/He$_\odot$ & 1.26  & 1.38  & 1.45  & 1.55  & 1.41  & 1.21  & 1.17  & $\dots$ \\
N/O           & 0.31  & 0.29  & 0.58  & 0.28  & 0.46  & 0.32  & 0.43  & 0.13 \\
Ne/O          & 0.48  & 0.89  & 0.51  & 0.65  & 0.57  & 0.21  & 0.37  & 0.21 \\
Ar/O          & 0.05: & 0.04: & 0.05: & 0.03: & 0.04: & 0.02: & 0.02: & 0.006 \\
S/O           & 0.04  & 0.02  & 0.05  & 0.03  & 0.05  & 0.03  & 0.03  & 0.03 \\
\hline
\end{tabular}
\vspace{0.4cm}
\label{sample}
\end{table*}

\subsection{A comprehensive view of NGC\,2371}

The spatio-kinematic model of NGC\,2371 described above can be compared to its excitation structure and spatially varying chemical abundances. Despite its complex morphology, NGC\,2371 is quite symmetric, with an ellipsoidal main nebular shell with polar blowouts and a pair of bipolar lobes whose symmetry axis is aligned along a slightly tilted direction. The tips of the bipolar lobes show clear evidence of shock-excitation, which can arise as the nebular material expands at relatively high speed and interacts with the ISM. On the other hand, the main nebular shell and very notably the regions close to the CSPN exhibit very high excitation, with the lowest [N~{\sc ii}]/H$\alpha$ and highest He\,{\sc ii}/H$\beta$ line ratios. Indeed, excitation in these innermost regions is so high that the [O~{\sc iii}]/H$\alpha$ line ratio is damped as O$^{++}$ is photo-ionised to higher ionisation levels. Furthermore, several He\,{\sc ii} emission lines can be seen in the optical spectrum shown in Figure~5. Along with C~{\sc iv} and [Ne~{\sc iv}], it has also been reported to be present in the \emph{IUE} UV nebular spectrum of NGC\,2371 \citep[see fig.~2 in][]{Pottasch1981}. Since the He~{\sc ii} $\lambda$1640 \AA\ is the dominant line in the wavelength range between 1400 \AA\ and 1800 \AA, the \emph{GALEX} far-UV image in Figure~12 (left panel) can be used to probe the spatial location of high-excitation material in NGC\,2371. The emission in this image traces the main morphological features of NGC\,2371, including the main cavity, the dense knots, and the outer edges of the NW and SE lobes. In addition, there is a diffuse halo that extends to distances of $\sim70$~arcsec from the CSPN, which is not detected in the near-UV {\it GALEX} image (see Fig.~12 right panel). When comparing with mid-IR \emph{Spitzer} presented in \citet{RamosLarios2012} and optical CFHT [O~{\sc iii}] images, there is notable emission in the \emph{Spitzer} images close to the CSPN, most likely associated with high-excitation [O~{\sc iv}] emission lines. The high excitation of the inner regions of NGC\,2371 results from the strong UV flux of WD\,0722$+$295, whereas the high-excitation of the outer halo might result from the "hardening" of the stellar radiation, as high energy photons leak from the main nebular shell of NGC\,2371 \citep[see][]{Guerrero1999,Gesicki2003}. 

The low-ionisation dense SW and NE knots are aligned along a direction different from the main nebular axis and they do not share the expansion velocity of the main nebula, with lower expansion velocities ($\pm$7~km~s$^{-1}$).  
Their morphology, with bow-shocks pointing towards the ionisation source, 
also precludes a jet interpretation for these features, but they are more 
consistent with clumps surrounded by photo-evaporated material.  
Thus, whereas the main nebula of NGC\,2371 has very high-excitation, it 
seems that the face of these knots pointing towards the CSPN shields their cometary tails
from the UV stellar flux, producing emission from low ionisation 
species at their tails. 
Indeed, the spectra extracted from these structures present emission from [O\,{\sc i}] and [N\,{\sc i}] (see A2, A7 and A8 in Tab.~2) suggesting them 
to be low-ionisation structures (LIS).
The analysis of LIS in PNe presented by \citet{Akras2016} suggests
$T_\mathrm{e}$=10,000--14,700~K, consistent with our estimates for 
the SW and NE knots, but lower electron densities than the surrounding 
ionised gas, contrary to the SW and NE knots of NGC\,2371, which are 
much denser than the nebular material where they are embedded (see Tab.~2). 
In this sense, the SW and NE knots of NGC\,2371 are not typical LIS.

The chemical abundances of different apertures along the minor axis 
of NGC\,2371 are listed in Table~5.  
The mean value of the He abundances is 0.113$\pm$0.012 
\cite[$\simeq$1.4 times the solar value in][]{Lodders2010}, 
with most individual values within 1-$\sigma$ uncertainty.  
This might also be the case for the O abundances, with a mean value of  
(4.1$\pm$2.3)$\times$10$^{-4}$, which is mostly consistent within 1-$\sigma$ 
uncertainty with that of individual apertures.  
This mean value suggests slightly subsolar O abundances, 
(O/H)$\simeq$0.75 (O/H)$_\odot$. 
On the contrary, the N abundances are clearly larger than solar, 
with a mean N/O ratio of 0.38$\pm$0.11, i.e., (N/O)$\approx$3 
(N/O)$_\odot$.  
The chemical abundances (He/H, O/H, N/O) of NGC\,2371 are basically 
consistent with those reported for comprehensive samples of PNe with 
[WC] central stars \citep{Pena2001,GarciaRojas2013}.  
The other chemical abundances in Table~5 have larger uncertainties, 
particularly the Ar abundances, for which no error bar is provided.  
The values of the chemical abundances of Ne, Ar, S, and Cl are 
also consistent with those of PNe with [WC] central stars.  

The chemical abundances found in different regions are in general consistent 
among them, 
within 2$\sigma$ up most from their mean values,
but we note that the abundances
of the main elements He, N, and O show some
differences from 
one region to another.  
It might be argued that the O abundances from apertures probing the 
dense low-ionisation structures are higher than those of the main cavity enclosing them, whereas the N/O ratios are lower. 
We believe this is not the case, because a
close inspection reveals that the O abundances are mostly 
anti-correlated with $T_\mathrm{e}$, underlying the difficulties to 
determine the chemical abundances in highly excited nebulae and from 
regions where low- and high-excitation material coexists.

Finally, we would like to remark that \citet{Herald2004} suggested  
WD\,0722$+$295 to be in the same evolutionary stage as the CSPN of 
the born-again PN A\,78, a [WR]-PG 1159 star.  
Although their analysis of the stellar atmospheres of those two CSPNe with 
their version of the CMFGEN code resulted in similar atmospheres, our 
best-fit models of WD\,0722$+$295 and our recent analysis of the CSPN of 
A\,78 \citep[see][]{Toala2015} show these two stars to be different.  
Moreover, the presence of hydrogen-deficient material inside 
A\,78 suggests a different evolutionary path compared to the 
CSPN of NGC\,2371. 
The spectral analysis of WD\,0722$+$295 presented here represents an 
improvement over other similar studies due to the high-quality of the 
optical spectrum and the combination with available UV data used for 
the fit. 

The PoWR model presented here is not able to reproduce the lines identified 
in Section~3 at $\approx$5665 \AA\ and $\approx$6066 \AA\ as O\,{\sc vii} 
$\lambda$5666 and O\,{\sc viii} $\lambda$6068, because the ionisation 
potentials of these species, 739.3~eV and 871.4~eV, respectively, are too 
high to be produced by the stellar atmosphere.
Although it can be argued that such high-temperature gas can be produced by the X-ray emission of WD\,0722$+$296 detected by {\it Chandra} \citep[][]{Montez2015}, it has been long discussed whether these emission lines are due to oxygen, but to neon. 
\citet{Werner2007} presented stellar atmosphere models of pre-WD stars and 
demonstrated that most of the emission lines from the O\,{\sc vii} and 
O\,{\sc viii} high-ionisation states of oxygen might be misidentified with 
those of Ne\,{\sc vii} and Ne\,{\sc viii}, 
whose ionisation potentials, 207.3~eV and 239.1~eV, respectively, 
are considerably lower. 
This does not affect the spectral classification of WD\,0722$+$296 as 
part of the [WO] sequence, because the O\,{\sc vi} $\lambda$3820, 
O\,{\sc vi} $\lambda$5290, and O\,{\sc vi} $\lambda$6200 lines are still 
present in its spectrum (see Table~1). 
Indeed, the Ne\,{\sc vii} and Ne\,{\sc viii} lines require a high 
temperature for the central star.

\section{Summary}

We presented spatially-resolved longslit INT IDS spectroscopic observations of NGC\,2371, a PN with a [WR]-type CSPN with an apparent complex morphology. 
Our spectral observations, in conjunction with \emph{HST}, {\it GALEX} and CFHT images, have allowed us to characterise the central star of NGC\,2371 as well as to unveil its high-ionisation structure. Our findings can be summarised as follows:
\begin{itemize}
    
    \item We studied the spectral properties of the CSPN of NGC\,2371, WD\,0722$+$295, by analysing the IDS observations. A Gaussian-fitting procedure was performed to measure the true contribution of the WR spectral lines and to assess the spectral sub-type of the CSPN. Line ratios of WR features with that of the RB suggest a [WO1]. In combination with the UV data we used the stellar atmosphere code PoWR to estimate the stellar parameters. Although we found very similar results as those obtained in previous works, the differences are due to the assumed clumping factor and a more accurate {\it Gaia} distance.
    
    \item We studied the physical properties of different regions of NGC\,2371. In accordance with previous works on NGC\,2371 the electron density ranges between 100~cm$^{-3}$ and 1,640~cm$^{-4}$. The densest regions are the SW and NE low-ionisation knots. The electron temperatures vary from $T_\mathrm{e}\sim$12,000~K at the outer edges to $T_\mathrm{e}\approx18,000$~K for regions close to the star. The outer bipolar (NW and SE) lobes have electron densities $n_\mathrm{e}\approx400$~cm$^{-3}$ and $T_\mathrm{e}\approx12,000-13,000$~K. 
    
    \item The analysis of emission line ratios from the IDS spectra as well as those extracted from the {\it HST} calibrated images demonstrate the powerful effect of the UV flux from the CSPN of NGC\,2371. 
    Most emission from this PNe is dominated by ionisation, but the 
    low-ionisation SW and NE knots and the outer regions B1 and B9, 
    where shock-excitation is important. 
    We suggest that B1 and B9 are shock-excited because they trace the shock 
    due to the expansion of NGC\,2371. 
    
    \item Inspection of {\it GALEX} images unveiled the presence of a halo extended to 70~arcsec from WD\,0722$+$295. The halo is detected in the far-UV channel and is mainly dominated by He\,{\sc ii} emission. We suggest that this halo has formed due to high-energy photon leakage.
    
    \item 
    We found that the bow shock-like structure around the CSPN-facing head 
    of the low-ionisation dense knots correspond to photoevaporation flows. 
    The head of the knots shield their tails producing emission from low-ionisation species such as [O\,{\sc i}] and [N\,{\sc i}]. This reinforces the idea proposed by \citet{Herald2004} and \citet{RamosLarios2012} that molecular material might be present in the vicinity of the CSPN of NGC\,2371, probably in these structures similar to LIS in other PNe \citep[][]{Akras2017}. However, we note that these clumps do not share the same physical properties as classic LIS described in the literature.
    
    \item We estimated the chemical abundances for different regions of 
    NGC\,2371.
    The abundances are typical of PNe with [WC] central stars.  
    Although some variations may exist among different regions, the abundances 
    of the dense knots lay within those of adjacent regions, suggesting that 
    they were not ejected as the result of a VLTP.  
    Instead, their kinematics and detailed morphology suggest they were 
    ejected before the formation of the main nebular shell. 
    
    \item The relatively high-resolution of the optical longslit spectra presented here allowed us to suggest a possible kinematic structure of NGC\,2371. The outer lobes have the typical expansion velocity of evolved PNe ($\sim$30~km~s$^{-1}$) whilst the inner shell has an expansion velocity of $\sim$70~km~s$^{-1}$ in accordance with previous analysis of high-resolution echelle observations. On the other hand, the dense knots have a slow radial velocity. We suggest that NGC\,2371 has a bipolar shape with each lobe presenting a double-structure protruding from a barrel-like central region.
    
\end{itemize}

We propose that the densest material around WD\,0722$+$295 might be the
relic of an early ejection of material along a ring-like structure. 
The ejection of the main nebula is more recent and can
 be directly related to the current fast stellar wind of the progenitor star 
 of NGC\,2371. 
 It seems to have expanded towards the low-density region, the polar zone 
 of the barrel-like structure. 
 We emphasise that the analysis of high-resolution echelle observations 
 such as those obtained with the San Pedro M\'{a}rtir MES are most 
 needed in order to construct a detailed view of the morpho-kinematic 
 structure of NGC\,2371.  

\section*{Acknowledgements}

The authors thank the referee for a critical reading of the manuscript and valuable suggestions that improved the presentation of the paper. The authors would like to thank V.\,G\'{o}mez Llanos Sandoval for helping them in using PyNEB. VMAGG, JAT, MAG and HT are funded by UNAM DGAPA PAPIIT project
IA100318. GRL acknowledges support from Fundaci\'on Marcos Moshinsky, CONACyT and PRODEP (Mexico). 
MAG acknowledges support of the Spanish Ministerio de Ciencia, Innovación y Universidades grant PGC2018-102184-B-I00, co-funded by FEDER funds.
 LS thanks support from UNAM PAPIIT grant IN101819. YDM thanks CONACyT for the research grant CB-A1-S-25070. This work has make extensive use of the NASA's Astrophysics Data System. This paper also presents data obtained with the MOS camera at the Canada-France-Hawaii Telescope (CFHT) which is operated by the National Research Council (NRC) of Canada, the Institut National des Sciences de l'Univers of the Centre National de la Recherche Scientifique of France, and the University of Hawaii. The authors thank E.\,Santamar\'{i}a for helping produce the velocity sketch of NGC\,2371. Based on observations made with the NASA/ESA Hubble Space Telescope, obtained at the Space Telescope Science Institute, which is operated by the Association of Universities for Research in Astronomy, Inc., under NASA contract NAS 5-26555. 



%\bibliographystyle{mnras}


{99}

\end{document}